\documentclass[10pt,prc,showpacs,preprintnumbers,amsmath,amssymb,aps,nofootinbib]{revtex4-1}
\usepackage[utf8]{inputenc}
\usepackage{amsmath}
\usepackage{graphicx}
\usepackage{amsfonts}
\usepackage{amssymb}
\usepackage{color}
\usepackage{ulem}
\allowdisplaybreaks[1]
\graphicspath{{pics/}}

\begin{document}
\title{Influence of confining  gluon configurations on the $P\to\gamma^*\gamma $ transition form factors}
\author{ Sergei N.  Nedelko}
\email{nedelko@theor.jinr.ru}
\author{Vladimir E. Voronin}
\email{voronin@theor.jinr.ru}
\affiliation{ Bogoliubov Laboratory of Theoretical Physics, JINR,
141980 Dubna, Russia }
\begin{abstract}
Transition form factors  $F_{P\gamma^*\gamma^{(*)}}$ of pseudoscalar mesons are studied within the framework of the domain model of confinement, chiral symmetry breaking and hadronization. In this model, the QCD vacuum is described by the statistical ensemble of  domain wall networks  which  represents  the almost everywhere homogeneous Abelian (anti-)self-dual gluon field configurations. Calculations of the form factors are performed consistently with mass spectra of light, heavy-light and double-heavy mesons,  their weak  and strong decay constants.   Influence of the  nonperturbative intermediate  range gluon fields on asymptotic behaviour of pion transition form factor is of particular interest, as it can potentially lead to the growth of $Q^2F_{\pi\gamma^*\gamma}(Q^2)$.   It is found that $Q^2F_{\pi\gamma^*\gamma}(Q^2)$ approaches a constant value  at asymptotically large $Q^2$.   However, this limit differs from the standard factorization bound,   though for pion form factor  complies  with Belle data more likely than with BaBar ones. At the same time  the generally accepted factorization bound is shown to be  satisfied  for the case of the  symmetric kinematics,  $Q^2F_{P\gamma^*\gamma^*}(Q^2)$.  Peculiarities of description of  $\eta$, $\eta'$ and $\eta_c$   form factors  within the model are discussed in detail.
\end{abstract}
\maketitle

\section{Introduction}
Experimental data for pion transition form factor $F_{\pi\gamma^*\gamma}$ obtained by the BaBar collaboration~\cite{Aubert:2009mc} indicate growth of $Q^2 F_{\pi\gamma^*\gamma}$ at large $Q^2$ that is inconsistent with  prediction of QCD factorization theorems~\cite{Lepage:1980fj}
\begin{equation}
\label{form factor_factorization_asym}
F_{\pi\gamma^*\gamma}\sim \frac{\sqrt{2}f_\pi}{3Q^2}\int_0^1 dx \frac{\phi^\mathrm{as}_\pi(x)}{x}= \frac{\sqrt{2} f_\pi}{Q^2},\quad \phi^\mathrm{as}_\pi(x)=6x(1-x),
\end{equation}
where $f_\pi=131\ \mathrm{MeV}$, and $\phi^\mathrm{as}_\pi(x)$ is the asymptotics of pion distribution amplitude at large $Q^2$. Published later experimental results  carried out by the Belle~\cite{Uehara:2012ag} collaboration demonstrate qualitatively different behavior at large momenta, though they still allow violation of bound~\eqref{form factor_factorization_asym}. These intriguing though so far  incomplete experimental results  have motivated extensive theoretical investigations of pion electromagnetic transition form factor $F_{\pi\gamma^*\gamma}$ and QCD factorization in exclusive hadronic processes. 
Transition form factors were considered within the framework of light-cone~\cite{Bakulev:2012nh,Mikhailov:2009kf,Agaev:2010aq} and anomaly sum rules~\cite{Klopot:2012hd,Oganesian:2015ucv}, local-duality version of QCD sum rules~\cite{Lucha:2011if}, modified perturbative approach based on the $k_T$ factorization theorem~\cite{Li:2009pr,Kroll:2010bf},  dispersion relations~\cite{Gorchtein:2011vf}, light-front holographic QCD~\cite{Brodsky:2011yv,Swarnkar:2015osa,Zuo:2009hz,Stoffers:2011xe},  Dyson-Schwinger equations~\cite{Roberts:2010rn}, nonlocal chiral quark models~\cite{Dorokhov:2010bz,Dorokhov:2010zzb, Dorokhov:2013xpa,Kotko:2009mb}, light-front quark model~\cite{Lih:2012yu}, vector-meson dominance model and its modifications~\cite{Lichard:2010ap,Arriola:2010aq}, within chiral effective theory with resonances~\cite{Czyz:2012nq}, instanton liquid model~\cite{Kochelev:2009nz}, models involving physics beyond the Standard Model~\cite{McKeen:2011aa}.
 Contribution of Adler-Bell-Jackiw triangle anomaly to $\pi^0\to\gamma^* \gamma $ process was investigated in~\cite{Pham:2011zi}.

Disagreement between bound~\eqref{form factor_factorization_asym} and the BaBar data aroused discussion about validity of the latter~\cite{Mikhailov:2009kf,Roberts:2010rn}. The findings demonstrate that higher local operator product expansion  and $\alpha_s$ corrections to Eq.~\eqref{form factor_factorization_asym} are too small to describe BaBar data, and one should take into account non-local or non-OPE contributions~\cite{Klopot:2012hd}. These corrections may originate from nonlocal condensates, instantons, or short strings. Alternatively, growth of $Q^2 F_{\pi\gamma^*\gamma}$ at large $Q^2$ can be described by "flat" (non-vanishing at the endpoints $x=0$ and $x=1$) distribution amplitude~\cite{Dorokhov:2009dg,Radyushkin:2009zg,Polyakov:2009je}.

\begin{table}
\caption{Values of parameters fitted to the mass spectrum~\cite{Nedelko:2016gdk} and used for calculations in the present paper.}
{\begin{tabular}{@{}ccccccc@{}}
 \toprule
$m_{u/d}$(MeV)&$m_s$(MeV)&$m_c$(MeV)&$m_b$(MeV)&$\Lambda$(MeV)&$\alpha_s$&$R$(fm)\\
\colrule
$145$&$376$&$1566$&$4879$&$416$&$3.45$&$1.12$\\
\botrule
\end{tabular}
\label{values_of_parameters}}
\end{table}

In this paper we consider behaviour of the transition  form factors within the approach based on the description of QCD vacuum as statistical ensemble of domain wall networks representing an ensemble of almost everywhere homogeneous Abelian (anti-)self-dual fields, which are characterized by the nonzero gluon  condensates, first of all the  scalar $\langle g^2F^2 \rangle$  and the absolute value of the pseudoscalar $\langle |g^2\tilde F F| \rangle=\langle g^2F^2 \rangle $ ones.   Motivation for this approach as well as  details related to the study of static and dynamical  confinement, realization of chiral $SU_{L}(N_f)\times SU_{R}(N_f)$ and $U_A(1)$ symmetries in terms of quark-gluon as well as colorless hadron degrees of freedom  can be found in papers~\cite{EN1,EN,NK1,NK4,Nedelko:2014sla,Nedelko:2016gdk} and references therein. The effective meson action derived in the model allows one to compute the mass spectrum,   decay and transition constants as well as form factors describing the strong, electromagnetic and weak interactions  of various mesons.
Parameters of the model are the infra-red limits of the renormalized quark masses and strong coupling constant, scalar gluon condensate $\langle \alpha_s F^2 \rangle$ and the mean domain size. The latter can be related to topological susceptibility of pure Yang-Mills vacuum. 
Overall precision of the approach  in description of wide range of meson phenomenology (masses of light, heavy-light  mesons and heavy quarkonia, leptonic decay constants, transition constants, all of the above including excited mesons)   is about $10-15\%$ with  few exceptions. Throughout all calculations the same values of parameters are used as it is supposed by their physical meaning~\cite{Nedelko:2016gdk}, see Table~\ref{values_of_parameters}. 

The main purpose of the present paper is to investigate how the explicit presence of the background domain structured gluon field  affects transition form factors of pseudoscalar mesons and strong decay constants of vector mesons $g_{VPP}$.  In this context the most relevant feature of the present approach is  the invariance of meson effective action with respect to the local gauge transformations of the background field. The nonlocal meson-quark vertices depend on the covariant derivatives in the presence of nonperturbative domain structured gluon field. Within the formalism used in this paper both quark propagators and meson-quark vertices are translation invariant up to a gauge transformation.  As a result, energy-momentum is conserved only in the entire diagram describing interaction between mesons, leptons and photons, but it is not conserved in every meson-quark vertex separately. 
Averaging of these diagrams over configurations of the nonperturbative field leads to contributions that are absent if just a global gauge invariance is assumed as it is usually done in nonlocal models of hadronization.
 This is the most relevant to our task  feature of the effective meson action, derived within the domain model, as it  becomes a source for violation of factorization properties since it may mix up  soft and hard parts of the amplitudes,  short and large distance subprocesses.  It is shown that the  background gluon fields, typical for the domain model, do not cause growth of $Q^2 F_{\pi\gamma^*\gamma}$ at large $Q^2$, and calculation of the asymptotic behaviour indicates   that
\begin{equation}
\label{pigammagamma*}
F_{\pi\gamma^*\gamma}\sim \varkappa_{\gamma^*\gamma}\frac{\sqrt{2}f_\pi}{Q^2}, \ \varkappa_{\gamma^*\gamma}=1.23,
\end{equation}
 so that $Q^2 F_{\pi\gamma^*\gamma}$ approaches a constant value at large $Q^2$ in qualitative agreement with factorization prediction, but the value of constant $\varkappa_{\gamma^*\gamma}$ substantially differs from unity. At the same time, asymptotic behavior of form factor in symmetric kinematics with two photons with equal virtuality $Q^2$,
\begin{equation}
\label{pigamma*gamma*}
F_{\pi\gamma^*\gamma^*}\sim \varkappa_{\gamma^*\gamma^*}\frac{\sqrt{2}f_\pi}{3Q^2},
\ \varkappa_{\gamma^*\gamma^*}=1,
\end{equation}
calculated within the present model matches factorization prediction
\begin{equation}
\label{form factor_factorization_sym}
F_{\pi\gamma^*\gamma^*}\sim \frac{\sqrt{2}f_\pi}{3Q^2}\int_0^1 dx\ \phi^\mathrm{as}_\pi(x)= \frac{\sqrt{2} f_\pi}{3Q^2}.
\end{equation}
Within the present calculation the deviation of  $\varkappa_{\gamma^*\gamma}$ from unity  manifestly originates from  the non-conservation of the energy-momentum in the quark-meson vertices and quark propagators  separately due to  interaction of quarks and gluons with the background confining gluon fields. The energy-momentum is conserved  only for the whole locally gauge invariant amplitude of the process $\pi\to \gamma^*\gamma^{(*)}$. In the presence of the vacuum gluon fields under consideration  the propagators and vertices are translation invariant only up to a  gauge transformation of the background gluon fields. Equivalently,  in the presence  of the long/intermediate range vacuum gluon fields the locally gauge invariant amplitudes built of the nonperturbative quark propagators and vertices  are translation invariant  and, equivalently, ensure energy-momentum conservation.  

For symmetric kinematics these specific effects are suppressed and do not influence the asymptotic behaviour, while they directly contribute to asymptotic behavior of the form factor for asymmetric kinematics. 
  Relation~\eqref{form factor_factorization_sym} is reproduced much better than~\eqref{form factor_factorization_asym}.  This observation  looks  natural because a straightforward QCD factorization works best of all in the former case, i.e. when both photons are highly virtual~\cite{Radyushkin:1996tb,Mikhailov:2009kf}.

It is important for overall consistency of the formalism  that analogous terms originating from the local gauge invariance of the physical amplitudes critically affect the strong decays of vector mesons into a couple of pseudoscalar ones. Relaxation of the local   color gauge invariance to the global one leads to   drastic disagreement  between experimental and calculated  values of the  decay constants $g_{VPP}$,  particularly $g_{\rho\pi\pi}$. In this context,  strong underestimation of decay constant $g_{\rho\pi\pi}$ typical for NJL-type models of  hadronization~\cite{Bernard:1993wf,Deng:2013uca} can not be attributed just to the oversimplified local character of meson-quark interaction, nonlocality itself is not sufficient for consistent description of masses and strong decay constants.  The  local color gauge invariance of  the nonlocal  effective  meson-meson interactions mediated by the  quark-gluon interactions appears to be of crucial importance.

The paper is organized as follows. Brief review of the effective meson action is given in Sect.~\ref{section_domain_model}. Details of calculations of form factors are presented in Sect.~\ref{section_form factors}. Section \ref{section_vpp_decays} is devoted to decay constants of vector mesons into a couple of pseudoscalar mesons. Section~\ref{section_conclusions} contains conclusions and discussion of the unsolved problems. Details of calculations are given in the appendices. 

\begin{figure}[h!]
\begin{center}
\includegraphics[scale=1]{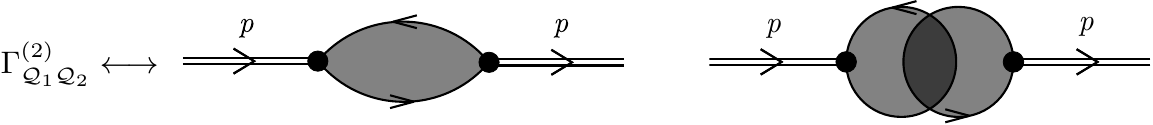}\end{center}
\caption{Diagrams contributing to the equation for the meson masses \eqref{mass-eq}  and quark-meson coupling constants \eqref{hqm}. Light grey color denotes averaging over configurations of the domain bulk background field. Dark grey color denotes  correlation of the quark loops directly due to the finite range correlators in the domain ensemble.  The two-loop diagram is particularly important for the masses of $\eta$ and $\eta^\prime$.
\label{diagrams}}
\end{figure}

\section{Effective meson action\label{section_domain_model}}
Euclidean functional integral of the domain model with the effective meson action  derived by means of hadronization procedure   has the following structure (for details of motivation and derivations see~\cite{Nedelko:2016gdk} and references therein):
\begin{gather}
Z={\cal N}
\int D\phi_{\cal Q}
\exp\left\{-\frac{\Lambda^2}{2}\frac{h^2_{\cal Q}}{g^2 C^2_\mathcal{Q}}\int d^4x 
\phi^2_{\cal Q}(x)
-\sum\limits_{k=2}^\infty\frac{1}{k}W_k[\phi]\right\},
\label{meson_pf}\\
W_k[\phi]=
\sum\limits_{{\cal Q}_1\dots{\cal Q}_k}h_{{\cal Q}_1}\dots h_{{\cal Q}_k}
\int d^4x_1\dots\int d^4x_k
\Phi_{{\cal Q}_1}(x_1)\dots \Phi_{{\cal Q}_k}(x_k)
\Gamma^{(k)}_{{\cal Q}_1\dots{\cal Q}_k}(x_1,\dots,x_k),
\label{effective_meson_action}
\\
\Phi_{{\cal Q}}(x)=\int \frac{d^4p}{(2\pi)^4}e^{ipx}{\mathcal O}_{{\mathcal Q}{\mathcal Q}'}(p)\tilde\phi_{{\mathcal Q}'}(p),\quad C_\mathcal{Q}=C_J,\quad C^2_{S/P}=2C^2_{V/A}=\frac{1}{9},
 \label{Pphi}\nonumber
\end{gather}
where condensed index $\mathcal{Q}\equiv\{aJLn\}$ denotes all meson quantum numbers like spin-parity in the ground state $J\in\{S,P,V,A\}$,  orbital momentum $l$ contributing to the  total  angular momentum (spin) for orbital excitations, radial quantum number $n$, flavour $SU(N)$ multiplets $a$ and space-time indices.
The meson masses $M_\mathcal{Q}$   and quark-meson coupling constants  $h_{\cal Q}$ are determined by the quadratic part of the effective meson action \textit{via} equations
\begin{gather}
\label{mass-eq}
1=
\frac{g^2 C^2_\mathcal{Q}}{\Lambda^2}\tilde \Pi_{\cal Q}(-M^2_{\cal Q}|B),
\\
h^{-2}_{\cal Q}=
\frac{d}{dp^2}\tilde\Pi_{\cal Q}(p^2)|_{p^2=-M^2_{\cal Q}}, 
\label{hqm}
\end{gather}
where $\tilde\Pi_{\cal Q}(p^2)$ is the diagonalized with respect to all quantum numbers two-point correlator $\tilde\Gamma^{(2)}_{\cal QQ'}(p)$  put on mass shell $p^2=-M^2_{\mathcal Q}$:
\begin{equation}
\label{diagonalization}
\tilde\phi^\dagger_{\mathcal Q}(-p)\left[\mathcal{O}^T(p)\tilde\Gamma^{(2)}(p)\mathcal{O}(p)\right]_{\cal QQ'}\tilde\phi_{{\mathcal Q}'}(p)|_{p^2=-M^2_{\cal Q}}=\tilde\Pi_{\cal Q}(-M_{\mathcal{Q}}^2)\tilde\phi^\dagger_{\mathcal Q}(-p)\tilde\phi_{\mathcal Q}(p)|_{p^2=-M^2_{\cal Q}}.
\end{equation}

Masses $M_{\mathcal Q}$ correspond to the poles of meson propagators, while relation \eqref{hqm} for meson-quark interaction constants  provides correct residue at the poles. 
This relation is known also as a  compositeness  condition for meson fields.
Integration variables $\phi_{\mathcal Q}$ in the functional integral \eqref{meson_pf} correspond to the physical meson fields that diagonalize the quadratic part of the effective meson action \eqref{effective_meson_action} in momentum representation, which is achieved by means of  transformation ${\mathcal O}(p)$.
Effective action \eqref{effective_meson_action} is expressed in terms of colorless composite fields $\Phi_Q$  related to the physical meson fields through the transformation $\mathcal{O}(p)$,
\begin{eqnarray*}
\tilde\Phi_{\mathcal{Q}}(p)={\mathcal O}_{\mathcal{Q}\mathcal{Q}'}(p)\tilde\phi_{\mathcal{Q}'}(p),
\end{eqnarray*}
 and  $k$-point nonlocal vertex functions $\Gamma^{(k)}_{\mathcal{Q}_1\dots \mathcal{Q}_k}$:
\begin{gather*}
\Gamma^{(2)}_{{\cal Q}_1{\cal Q}_2}=
\overline{G^{(2)}_{{\cal Q}_1{\cal Q}_2}(x_1,x_2)}-
\Xi_2(x_1-x_2)\overline{G^{(1)}_{{\cal Q}_1}G^{(1)}_{{\cal Q}_2}},
\\\nonumber
\Gamma^{(3)}_{{\cal Q}_1{\cal Q}_2{\cal Q}_3}=
\overline{G^{(3)}_{{\cal Q}_1{\cal Q}_2{\cal Q}_3}(x_1,x_2,x_3)}-
\frac{3}{2}\Xi_2(x_1-x_3)
\overline{G^{(2)}_{{\cal Q}_1{\cal Q}_2}(x_1,x_2)
G^{(1)}_{{\cal Q}_3}(x_3)}
\\
+
\frac{1}{2}\Xi_3(x_1,x_2,x_3)
\overline{G^{(1)}_{{\cal Q}_1}(x_1)G^{(1)}_{{\cal Q}_2}(x_2)
G^{(1)}_{{\cal Q}_3}(x_3)},
\\
\Gamma^{(4)}_{{\cal Q}_1{\cal Q}_2{\cal Q}_3{\cal Q}_4}=
\overline{G^{(4)}_{{\cal Q}_1{\cal Q}_2{\cal Q}_3{\cal Q}_4}
(x_1,x_2,x_3,x_4)}-
\frac{4}{3}\Xi_2(x_1-x_2)
\overline{G^{(1)}_{{\cal Q}_1}(x_1)
G^{(3)}_{{\cal Q}_2{\cal Q}_3{\cal Q}_4}(x_2,x_3,x_4)}
\nonumber\\\nonumber
-
\frac{1}{2}\Xi_2(x_1-x_3)
\overline{G^{(2)}_{{\cal Q}_1{\cal Q}_2}(x_1,x_2)
G^{(2)}_{{\cal Q}_3{\cal Q}_4}(x_3,x_4)}
\\\nonumber
+
\Xi_3(x_1,x_2,x_3)
\overline{G^{(1)}_{{\cal Q}_1}(x_1)G^{(1)}_{{\cal Q}_2}(x_2)
G^{(2)}_{{\cal Q}_3{\cal Q}_4}(x_3,x_4)}
\\
-\frac{1}{6}
\Xi_4(x_1,x_2,x_3,x_4)
\overline{G^{(1)}_{{\cal Q}_1}(x_1)G^{(1)}_{{\cal Q}_2}(x_2)
G^{(1)}_{{\cal Q}_3}(x_3)G^{(1)}_{{\cal Q}_4}(x_4)}.
\end{gather*}
The vertices $\Gamma^{(n)}$ are expressed \textit{via} quark loops $G^{(l)}_{{\cal Q}_1\dots{\cal Q}_k}$ averaged over the background field
\begin{eqnarray}
\label{barG}
&&\overline{G^{(k)}_{{\cal Q}_1\dots{\cal Q}_k}(x_1,\dots,x_k)}
=\int d \sigma_B
{\rm Tr}V_{{\cal Q}_1}\left(x_1|B\right)S\left(x_1,x_2|B\right)\dots
V_{{\cal Q}_k}\left(x_k|B\right)S\left(x_k,x_1|B\right),
\\
\nonumber
&&\overline{G^{(l)}_{{\cal Q}_1\dots{\cal Q}_l}(x_1,\dots,x_l)
G^{(k)}_{{\cal Q}_{l+1}\dots{\cal Q}_k}(x_{l+1},\dots,x_k)}
=
\\
\nonumber
&&\int d \sigma_B
{\rm Tr}\left\{
V_{{\cal Q}_1}\left(x_1|B\right)S\left(x_1,x_2|B\right)\dots
V_{{\cal Q}_k}\left(x_l|B\right)S\left(x_l,x_1|B\right)
\right\}\times
\\
\nonumber
&&
{\rm Tr}\left\{
V_{{\cal Q}_{l+1}}\left(x_{l+1}|B\right)S\left(x_{l+1},x_{l+2}|B\right)\dots
V_{{\cal Q}_k}\left(x_k|B\right)S\left(x_k,x_{l+1}|B\right)
\right\},
\end{eqnarray}
where bar denotes averaging over all configurations of the background gluon field with measure $d\sigma_B$. There are two types of contributions to vertex functions $\Gamma^{(k)}$  --  one-loop diagrams and $n\le k$-loop diagrams connected by the  correlators of the statistical ensemble of the nonperturbative background fields -- almost everywhere homogeneous (anti-)self-dual Abelian gluon fields.        

\begin{figure}[h]
\centering
\includegraphics[width=10cm,clip]{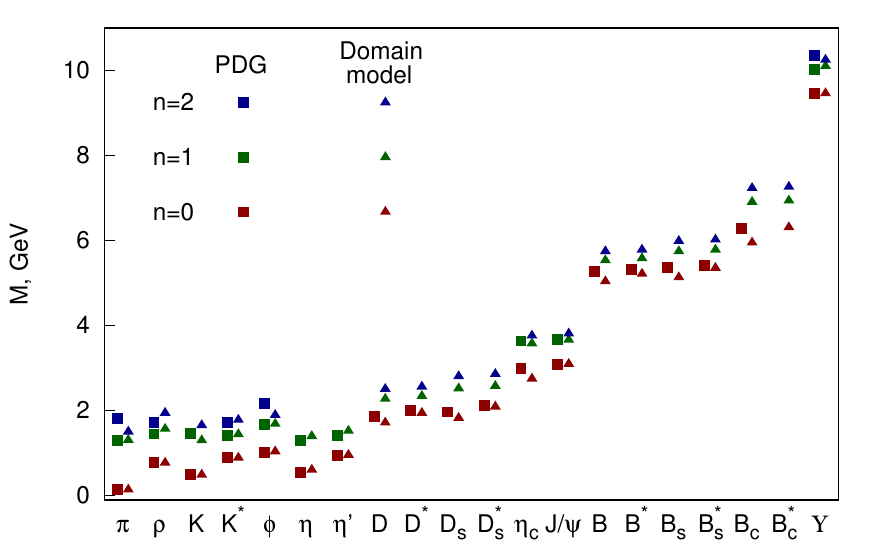}
\caption{The masses of various radially excited mesons calculated with the  values of parameters  shown in Table~\ref{fig-masses}. }
\label{fig-masses}       
\end{figure}

\begin{table}[ph]
\caption{Decay and transition constants of various  mesons \cite{Nedelko:2016gdk}. }
{\begin{tabular}{@{}cccc|cccc@{}} \toprule
Meson&$n$&$f_P^{\rm exp}$&$f_P$&Meson&$n$&$g^{\rm exp}_{V\gamma}$ 
\cite{PDG}&$g_{V\gamma}$\\
&&(MeV)& (MeV)&&&\\
\colrule
$\pi$      &0 &130 \cite{PDG}        &140 & $\rho$&0&0.2&0.2 \\
$\pi(1300)$&1 &                      &29 & $\rho$&1&&0.053 \\
\colrule
$K$        &0 &156 \cite{PDG}        &175  & $\omega$&0&0.059&0.067\\
$K(1460)$  &1 &                      &27   & $\omega$&1&&0.018\\
\colrule
$D$        &0 &205 \cite{PDG}        &212  & $\phi$&0&0.074&0.071\\
$D$        &1 &                      &51   & $\phi$&1&&0.02\\
\colrule
$D_s$      &0 &258 \cite{PDG}        &274  & $J/\psi$&0&0.09&0.06\\
$D_s$      &1 &                      &57  & $J/\psi$&1&&0.015\\
\colrule
$B$        &0 &191 \cite{PDG}        &187  & $\Upsilon$&0&0.025&0.014\\
$B$        &1 &                      &55   & $\Upsilon$&1&&0.0019\\
\colrule
$B_s$      &0 &253 \cite{Chiu:2007bc}&248  &  & &\\
$B_s$      &1 &                      &68   &  & &\\
\colrule
$B_c$      &0 &489 \cite{Chiu:2007bc}&434  &  & &\\
$B_c$      &1 &                  &135  & &&\\
\botrule
\end{tabular}
\label{constants}}
\end{table}

A simplified version of the domain model which allows one to compute the effective action analytically is based on two main approximations.  Quark and gluon propagators are computed in the homogeneous background field, which  represents the domain bulk, while  the finite size of domains is neglected at this step. The finite mean size is taken into account through the domain ensemble correlators  $\Xi_n(x_1,\dots,x_n)$ which have geometrical interpretation in terms of  
 a volume of overlap of $n$ four-dimensional hyperspheres with radius $R$  and centres at the points $x_1,\dots,x_k$ according to the formalism derived in ~\cite{NK1,NK4}.  As it has already been mentioned, meson  vertices $\Gamma^{(k)}_{\mathcal{Q}_1\dots \mathcal{Q}_k}$ are averaged over all configurations of the homogeneous (anti-)self-dual Abelian gluon fields: (anti-)self-duality, color   and space-time orientation. The latter is achieved by means of  generating formula
\begin{equation}\label{averaging_over_vacuum_field}
\langle\exp(if_{\mu\nu}J_{\mu\nu})\rangle=\frac{\sin\sqrt{2\left(J_{\mu\nu}J_{\mu\nu}\pm J_{\mu\nu}\widetilde{J}_{\mu\nu}\right)}}{\sqrt{2\left(J_{\mu\nu}J_{\mu\nu}\pm J_{\mu\nu}\widetilde{J}_{\mu\nu}\right)}},\quad f_{\alpha\beta}=\frac{\hat{n}}{2\upsilon\Lambda^2}B_{\alpha\beta}, \quad  \upsilon=\mathrm{diag}\left(\frac16,\frac16,\frac13\right),
\end{equation}
where tensor $f_{\mu\nu}$ is related to the strength of the Abelian (anti-)self-dual background field
\begin{gather*}
\label{b-field}
\hat B_\mu=-\frac{1}{2}\hat n B_{\mu\nu}x_\nu, \ \hat n = t^3\cos\xi+t^8\sin\xi,
\\ \tilde{B}_{\mu\nu}=\frac12\epsilon_{\mu\nu\alpha\beta}B_{\alpha\beta}=\pm B_{\mu\nu}, \  \hat{B}_{\rho\mu}\hat{B}_{\rho\nu}=4\upsilon^2\Lambda^4\delta_{\mu\nu},\\
f_{\alpha\beta}=\frac{\hat{n}}{2\upsilon\Lambda^2}B_{\alpha\beta}, \  \upsilon=\mathrm{diag}\left(\frac16,\frac16,\frac13\right), \ f_{\mu\alpha}f_{\nu\alpha}=\delta_{\mu\nu},
\end{gather*}
 and $J_{\mu\nu}$ is arbitrary antisymmetric tensor.  For example, generating formula leads to
\begin{gather*}
\langle f_{\mu\nu} \rangle=0,\\  \langle f_{\mu\nu} f_{\alpha\beta} \rangle = \frac{1}{3} \left(\delta_{\mu\alpha}\delta_{\nu\beta} - \delta_{\mu\beta}\delta_{\nu\alpha} \pm \varepsilon_{\mu\nu\alpha\beta}\right).
\end{gather*}
In this approximation  vertex  $V^{aJln}_{\mu_1\dots\mu_l}$ is defined by the formulas
\begin{gather}
V^{aJln}_{\mu_1\dots\mu_l}= {\cal C}_{ln}\mathcal{M}^a\Gamma^J F_{nl}\left(\frac{\stackrel{\leftrightarrow}{\cal D}^2\!\!\!
(x)}{\Lambda^2}\right)T^{(l)}_{\mu_1\dots\mu_l}\left(\frac{1}{i}\frac{\stackrel{\leftrightarrow}{\cal D}\!(x)}{\Lambda}\right),
\label{qmvert}\\
{\cal C}^2_{ln}=\frac{l+1}{2^ln!(n+l)!},\quad F_{nl}(s)=s^n\int_0^1 dt t^{n+l} \exp(st),
\nonumber\\
{\stackrel{\leftrightarrow}{\mathcal{D}}}\vphantom{D}^{ff'}_{\mu}=\xi_f\stackrel{\leftarrow}{\mathcal{D}}_{\mu}-\ \xi_{f'}\stackrel{\rightarrow}{\mathcal{D}}_{\mu}, 
\ \
\stackrel{\leftarrow}{\mathcal{D}}_{\mu}\hspace*{-0.3em}(x)=\stackrel{\leftarrow}{\partial}_\mu+\ i\hat B_\mu(x),  \ \ 
\stackrel{\rightarrow}{\mathcal{D}}_{\mu}\hspace*{-0.3em}(x)=\stackrel{\rightarrow}{\partial}_\mu-\ i\hat B_\mu(x), 
\nonumber\\
\xi_f=\frac{m_{f'}}{m_f+m_{f'}},\ \xi_{f'}=\frac{m_{f}}{m_f+m_{f'}},
\nonumber
\end{gather}
where $\mathcal{M}^a$ is the flavour matrix, $\Gamma^J$ is a  Dirac matrix with $J\in\{S,P,V,A\}$, and  $x$ is the center of mass of the quarks with  flavours $f$ and $f^\prime$ entering  and outgoing the vertex in \eqref{barG}. The form of the radial part  $F_{nl}$ of the vertex is determined by the propagator of the gluon fluctuations  charged with respect to the Abelian background.
Quark propagator in the presence of the Abelian (anti-)self-dual homogeneous field  has the form
\begin{eqnarray}
\label{quark_propagator}
&&S_f(x,y|B)=\exp\left(-\frac{i}{2}\hat n x_\mu  B_{\mu\nu}y_\nu\right)H_f(x-y|B),
\\
&&\tilde H_f(p|B)=\frac{1}{2\upsilon \Lambda^2} \int_0^1 ds e^{(-p^2/2\upsilon \Lambda^2)s}\left(\frac{1-s}{1+s}\right)^{m_f^2/4\upsilon \Lambda^2}
\label{quark_propagator_trinv}\nonumber\\
&&\times \left[\vphantom{\frac{s}{1-s^2}}p_\alpha\gamma_\alpha\pm is\gamma_5\gamma_\alpha f_{\alpha\beta} p_\beta
+m_f\left(P_\pm+P_\mp\frac{1+s^2}{1-s^2}-\frac{i}{2}\gamma_\alpha f_{\alpha\beta}\gamma_\beta\frac{s}{1-s^2}\right)\right].
\nonumber
\end{eqnarray}

\begin{figure}
\parbox{0.24\textwidth}{\includegraphics[scale=0.35]{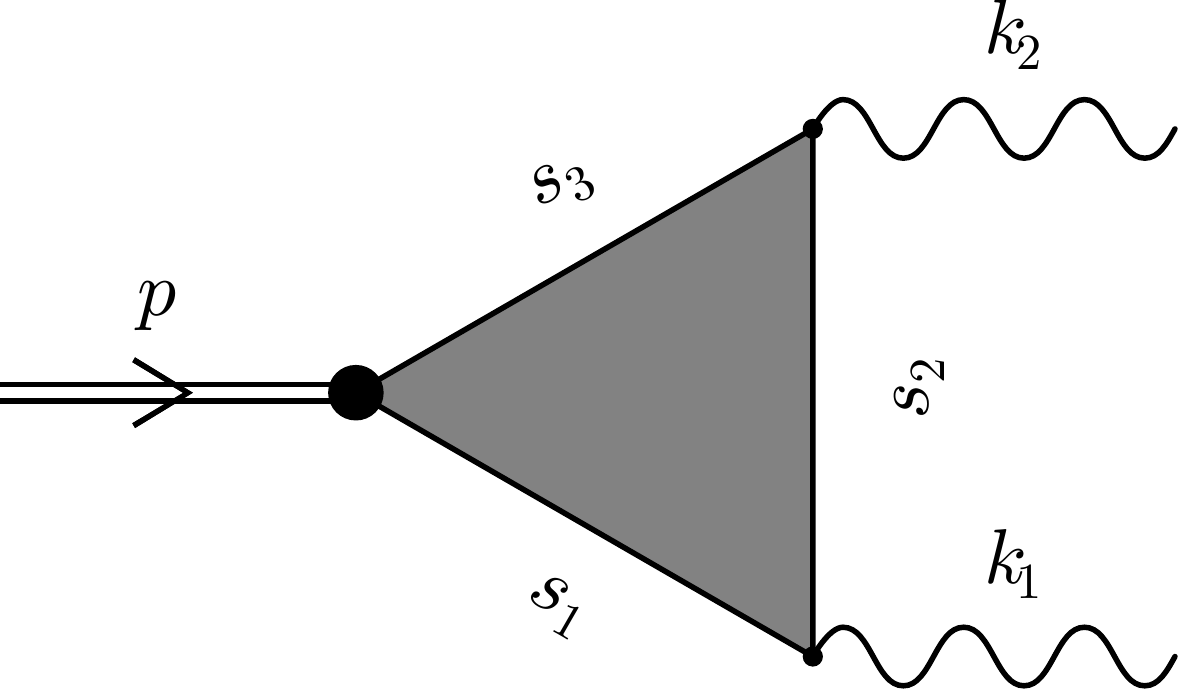}}
\parbox{0.2\textwidth}{\includegraphics[scale=0.35]{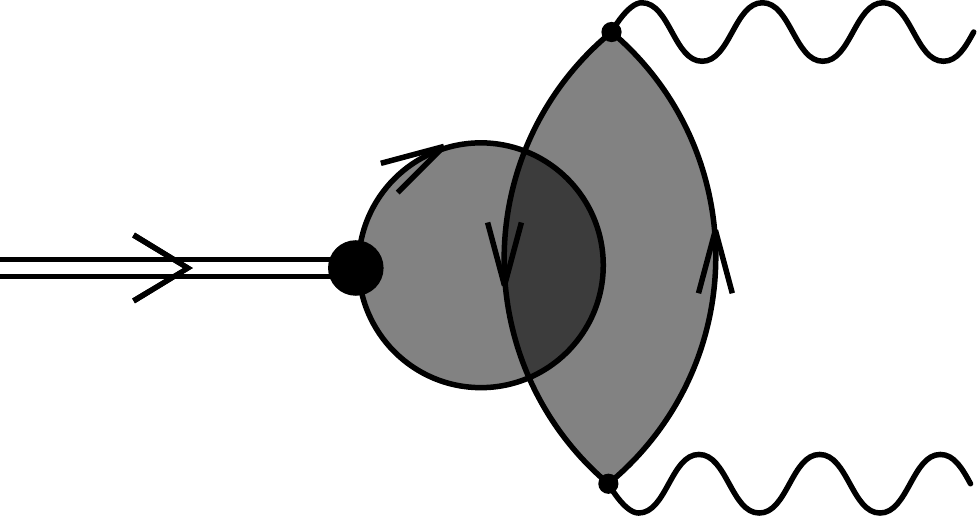}}
\parbox[c][\totalheight][t]{0.38\textwidth}{
\raisebox{-\height}{\includegraphics[scale=0.35]{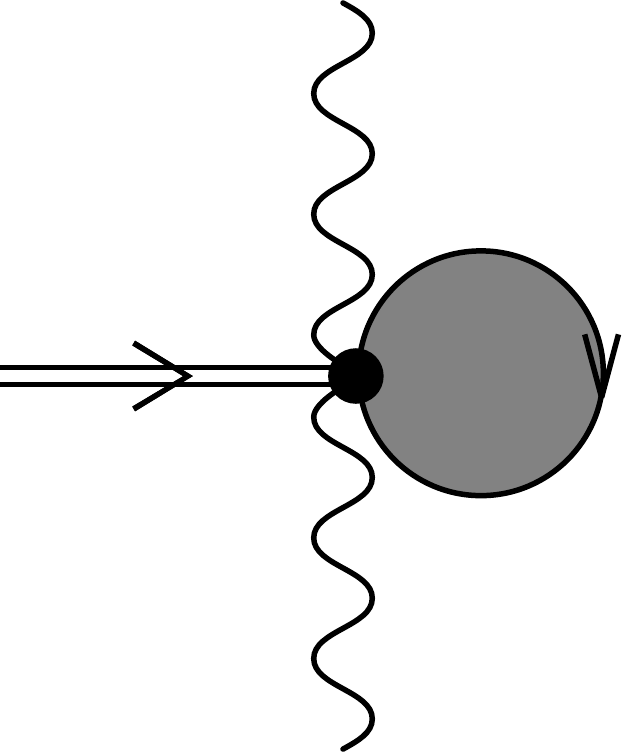}}
\hspace*{0.3em}
\raisebox{-\height}{\includegraphics[scale=0.35]{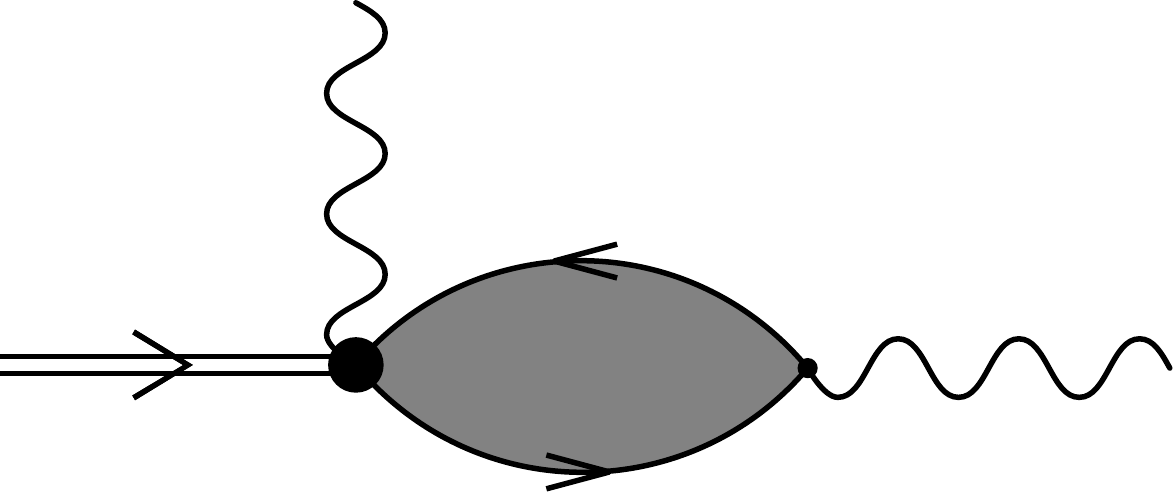}}
}\\[1em]
\parbox{0.24\textwidth}{A}
\parbox{0.2\textwidth}{B}
\parbox{0.13\textwidth}{C}
\parbox{0.24\textwidth}{D}
\caption{Diagrams which in principle could  contribute 
to the transition form factor.   In the approximation to the quark and gluon propagators that is used in the present calculation the only nonzero contribution comes from the diagram A. 
\label{F_P_gamma_picture}}
\end{figure}

\begin{table}
\caption{Matrix elements of the on-shell mixing matrix $\mathcal{O}_{\mathcal{Q}\mathcal{Q}^\prime}$
for $\pi$, $\eta$, $\eta^\prime$ and $\eta_c$. For $\pi$ and  $\eta_c$ mesons the quantities  $\mathcal{O}^{\pi/\eta_c}_{n}\equiv\mathcal{O}^{\pi/\eta_c}_{0n}(-M_{\pi/\eta_c}^2)$  characterize the weight  of the vertices \eqref{qmvert} with different radial number $n$ in interaction of quarks with the physical ground state $\pi$ and  $\eta_c$ mesons.  For $\eta$ and $\eta^\prime$ mesons mixing between octet and singlet states ($a=0,8$) is taken into account simultaneously with the radial number mixing.}
\begin{tabular}{ccccccccc}
\hline\hline
meson& a &$\mathcal{O}_{0}$& $\mathcal{O}_{1}$ &$\mathcal{O}_{2}$ & $\mathcal{O}_{3}$& $\mathcal{O}_{4}$ & $\mathcal{O}_{5}$& $\mathcal{O}_{6}$\\
\hline
$\pi$& -- &0.7595& -0.4510& 0.3067& -0.2294& 0.1826& -0.1515& 0.1293\\
$\eta_c$& -- &0.6225& -0.47789& 0.3788& -0.3079& 0.2554& -0.2155& 0.1846\\
\hline
$\eta$&0&0.244& -0.1437& 0.1036& -0.0812& 0.0662 &-0.0553& 0.0471\\
&8&-0.6724& 0.4495& -0.3189& 0.2406& -0.191& 0.1574 &-0.1334\\
\hline
$\eta'$&0&+0.0139& -0.5106& 0.4346& -0.2968& 0.198& -0.1388& 0.1049\\
&8&-0.6140 &-0.0201 &0.0985 &-0.0664 &0.0312 &-0.0096&-0.0013\\
\hline\hline
\end{tabular}\label{F_P_gamma_O}
\end{table}

\begin{table}\centering
\caption{Two-photon decay constants of  pseudoscalar mesons.
\label{p2gamma_constants}}
\begin{tabular}{c|c|c}
\toprule
meson\rule{0mm}{1.2em}&$g^{\rm exp}_{P\gamma\gamma},\ \mathrm{GeV}^{-1}$ \cite{PDG}&$g_{P\gamma\gamma},\ \mathrm{GeV}^{-1}$\\
\colrule
$\pi^0$    &0.274  &0.272\\
$\eta$     &0.274  &0.267\\
$\eta'$    &0.344  &0.44\\
$\eta_c$   &0.067  &0.055\\
\botrule
\end{tabular}\end{table}

The values of parameters of the model  given in Table~\ref{values_of_parameters} were fitted to the masses of ground-state mesons $\pi,\rho,K,K^*,J/\psi,\Upsilon,\eta'$ and were used for calculation of the decay and transition constants  as well as  masses   of a wide range of mesons~\cite{Nedelko:2016gdk}. The results are illustrated in  Fig.\ref{fig-masses} and Table~\ref{constants} respectively.

\section{Two-photon decay constants and transition form factors
\label{section_form factors}}

Weak and electromagnetic interactions can be introduced into meson effective action \eqref{effective_meson_action} by the standard requirement of the local gauge invariance which generates two types of gauge boson couplings -- the standard one with the local quark current  and direct interaction with the nonlocal quark-meson vertices. Detailed derivation for the model under consideration  can be found in paper~\cite{Nedelko:2016gdk}.   In particular, interaction of  the pseudoscalar mesons with two photons in the lowest order over quark-meson coupling constants (or, equivalently,  to the lowest order in $1/N_c$) is formally described by four terms represented diagrammatically in Fig.\ref{F_P_gamma_picture}, where  diagrams C and D include  additional meson-quark-photon coupling. However, contribution of diagrams C and D  vanish identically. Two other diagrams include the standard local electromagnetic interaction of quarks. Besides the  triangle diagram A there can be additional term B, which directly includes correlator of the nonperturbative gluon field. The latter term is analogous to the direct instanton contribution  considered in paper~\cite{Kochelev:2009nz} within the instanton liquid model.    A nonzero contribution to the form factors of $\pi$ and $\eta$ mesons  necessarily requires violation of  $SU(2)$  and/or  $SU(3)$ flavour symmetry respectively.   This condition is irrelevant to  $\eta^\prime$ and $\eta_c$  mesons, and one may expect that these mesons are much more sensitive to  correlations of the vacuum field ensemble crucial for  the two-loop diagram B. 
We shall return to discussion of this issue after studying the form factors generated by the 
 triangle diagram. Its contribution  to the  vertex  for interaction between the meson $\phi^{aP00}(x)$  and electromagnetic fields $A_\mu(y)$,  $A_\nu(z)$  has the form 
\begin{equation}
\label{triang}
T_{\mu\nu}^{an}(x,y,z)=h_{aP0n} \sum_{n^\prime,f} \mathcal{O}^{ab}_{nn^\prime} \mathrm{Tr}\int d\sigma_B   e_f^2\mathcal{M}_{ff}^b  F_{n^\prime 0} (x) i\gamma_5 S_f(x,y|B)\gamma_\mu S_f(y,z|B) \gamma_\nu S_f(z,x|B).
\end{equation}
Here  $f$ - flavour index, and  $\mathrm{Tr}$ denotes trace of the color and Dirac matrices. Quark-meson coupling constant $h_{aP00}$ is defined by \eqref{hqm}. Coefficients $\mathcal{O}^{ab}_{nn^\prime}$   describe mixing of the form factors $F_{n^\prime 0}$  in the quark-meson vertex for physical meson field put on shell, as it has come out of the   diagonalization of quadratic term \eqref{diagonalization} of the effective action and calculation of the  meson masses  \textit{via} Eq.\eqref{mass-eq}. For $\eta$ and $\eta'$ mesons $\mathcal{O}^{ab}_{0n}$ mixes not only the  vertices $F_{n0}$ but also  $\eta^0$ and  $\eta^8$ components of the pseudoscalar nonet  \eqref{triang} (see Table~\ref{F_P_gamma_O}). 
In momentum representation, the vertex  has the following structure:
\begin{equation}
T_{\mu\nu}^{an}(p^2,k_1^2,k_2^2)=ie^2\delta^{(4)}(p-k_1-k_2)\varepsilon_{\mu\nu\alpha\beta}k_{1\alpha}k_{2\beta}T^{an}(p^2,k_1^2,k_2^2).
\label{tmunu}
\end{equation}
Below we shall use dimensionless notation for the masses and momenta: $p^2\equiv p^2/\Lambda^2$, $k_1^2\equiv k_1^2/\Lambda^2$, $k_2^2\equiv k_2^2/\Lambda^2$,  $m_f\equiv m_f/\Lambda$.  Using  expressions ~\eqref{quark_propagator} and~\eqref{qmvert} for quark propagator and quark-meson vertices we arrive at 
\begin{eqnarray}
\label{formfactor_phases}
T^{an}(p^2,k_1^2,k_2^2)&=& 
\frac{h_{aP0n}}{16\pi^2\Lambda}\sum_{n^\prime,f,v}\frac{1}{v}\mathcal{O}^{ab}_{nn^\prime} \mathcal{M}_{ff}^b
q_f^2m_f  \int_0^1 ds_1 \int_0^1 ds_2 
\int_0^1 ds_3 \int_0^1 dt\ t^{n'} \frac{\partial^{n'}}{\partial t^{n'}}\\
&\times & 
\left[\left(\frac{1-s_1}{1+s_1}\right)\left(\frac{1-s_2}{1+s_2}\right) 
\left(\frac{1-s_3}{1+s_3}\right)\right]^{m_f^2/4v} 
\frac{1}{(1-s_1^2)(1-s_2^2)(1-s_3^2)}
\nonumber\\
&\times &\frac{1}{\phi^2}
\left[ \lambda_1 \frac{F_1}{\phi^2}+\lambda_2 
\left(\frac{F_2}{\phi^2}+m_f^2\frac{F_3}{\phi} -
p^2\frac{F_4}{\phi^3} 
+k_1^2\frac{F_5}{\phi^3}+k_2^2\frac{F_6}{\phi^3}\right)+\lambda_3 
\left[\left(p^2-k_1^2-k_2^2\right)^2-4k_1^2k_2^2\right]\frac{ 
F_7}{\phi^4} \right]
\nonumber\\
&\times & \exp \left\{-k_1^2\frac{ \phi_{11}}{2v\phi}-k_2^2\frac{\phi_{12}}{2v\phi}-p^2\frac{\phi_1}{4v\phi}\right\}.
 \nonumber
\end{eqnarray}
where $q_f$ is the electric charge of a quark with flavour $f$ in units of electron charge.   Polynomials $\phi,\phi_1,\phi_{11},\phi_{12}$  look as
\begin{gather}
\label{phi}
\phi = s_1+s_2 +s_3 + s_1s_2s_3 +2vt \left(1+s_1s_2 +s_1s_3+ 
s_2s_3\right), \\
\phi_1 =  \left(s_1- s_2+ s_3 
-s_1s_2s_3\right)vt +2s_1s_3,
\nonumber
\\
\phi_{11} =  s_2 [s_1+t v(1+s_1 s_3)  
],
\nonumber
\\
\phi_{12} =  s_2 [s_3+ tv(1+s_1 s_3)].
\nonumber
\end{gather}
Exact expressions  for  momentum independent polynomials   $F_i(s_1,s_2,s_3,t)$ are given in appendix \ref{transition_form factor_formula}.

\begin{figure}
\includegraphics[width=0.49\textwidth]{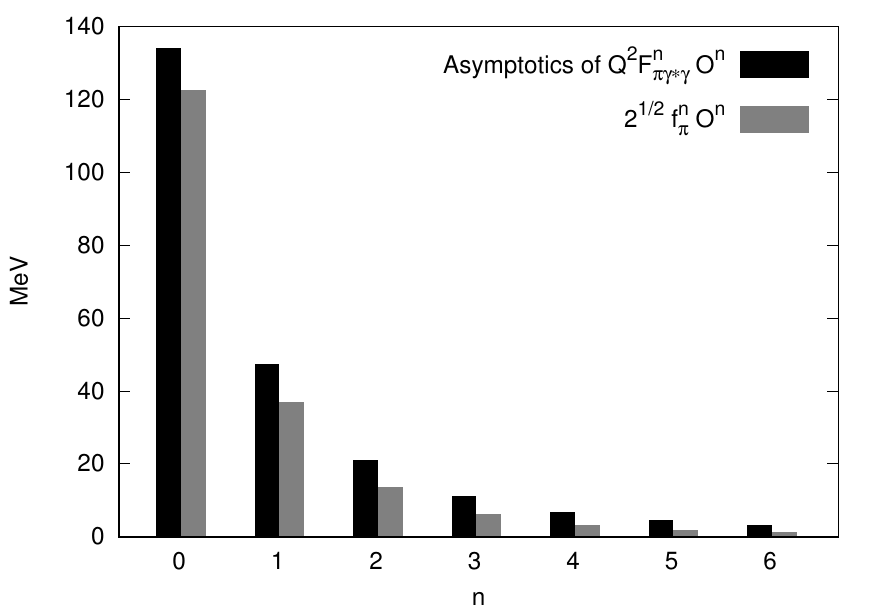}\hfill
\includegraphics[width=0.49\textwidth]{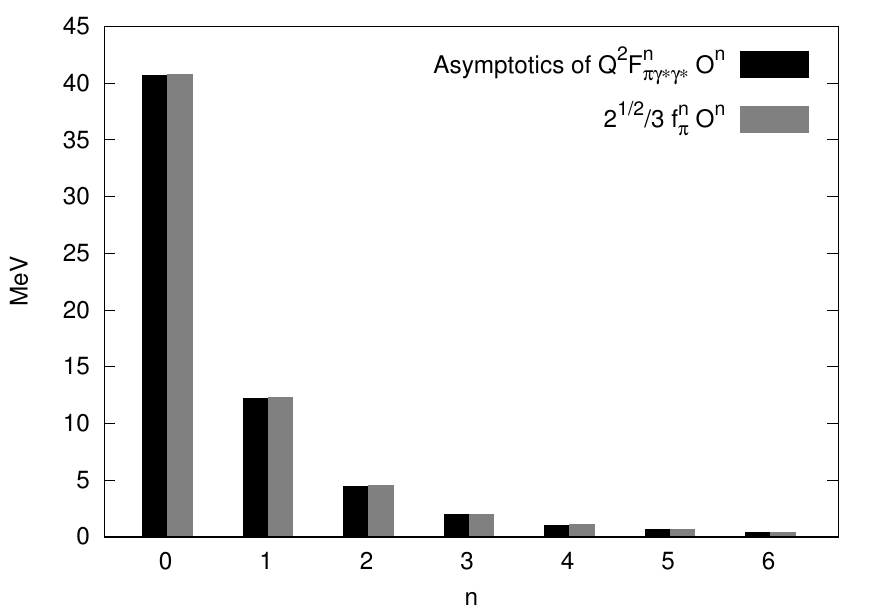}
\caption{Comparison of contributions to $\sqrt{2}f_\pi$ and large $Q^2$ asymptotics  of $Q^2F_{\pi\gamma^*\gamma}$ (lhs),   $\sqrt{2}f_\pi/3$ and asymptotics of $Q^2F_{\pi\gamma^*\gamma^*}$ (rhs)  from meson-quark vertices \eqref{qmvert} with various  radial  number $n$. For the sake of brevity  $O^n$ denotes  matrix elements $\mathcal{O}_{0n}^{33}$ which correspond to the ground state of $\pi^0$ that diagonalizes  quadratic part of the effective meson action.
The values of these matrix elements  are given in Table~\ref{F_P_gamma_O}.
\label{F_pigamma_versus_f_pi}}
\end{figure}

Functions $\lambda_i$ originate from averaging of the triangle diagram  over the space-time direction of the vacuum field according to Eq.~\eqref{averaging_over_vacuum_field}.  Before  averaging  the triangle diagram  is represented by the integral over  variables $s_1$, $s_2$, $s_3$ and $t$  with  integrand proportional to the exponential factor
\begin{eqnarray}
\label{phases}
&&\exp\left\{i f_{\mu\nu}J_{\mu\nu}\right\}, 
\nonumber\\
&& J_{\mu\nu}=\frac{\phi_2(s_1,s_2,s_3,t)}{2v\phi(s_1,s_2,s_3,t)}\left(k_{1\mu}k_{2\nu}-k_{1\nu}k_{2\mu}\right),
\\
&&\phi_2 = s_2 (s_1 s_3+  (s_1+s_3)t v).
\nonumber
\end{eqnarray}
This factor  is directly related to the  local gauge invariance of the meson effective action  with respect to the local gauge transformations of the background field which is provided by the covariant derivatives in the vertex \eqref{qmvert} and the exponential phase factor (sometimes called Schwinger phase)  in the quark propagator   \eqref{formfactor_phases}.
As a result of averaging the   final representation  contains functions 
\begin{eqnarray}
\label{f_i}
&&\lambda_1(r)=\frac{\sin r}{r}, \ 
\lambda_2(r)=-\frac{1}{r}\frac{\partial}{\partial r}\lambda_1(r), \ 
\lambda_3(r)=-\frac{1}{r}\frac{\partial}{\partial r}\lambda_2(r),
\\
&& r=\sqrt{2\left(J_{\mu\nu}J_{\mu\nu}\pm 
J_{\mu\nu}\widetilde{J}_{\mu\nu}\right)}=\frac{\phi_2}{v\phi}\sqrt{k_1^2k_2^2-\left(k_1k_2\right)^2}=\frac{\phi_2}{2v\phi}\sqrt{4k_1^2 
k_2^2 - (p^2-k_1^2-k_2^2)^2}.
\nonumber
\end{eqnarray}
For $p^2=-M^2$, that is relevant to the transition form factors,  the argument $r$ becomes imaginary $r=i\rho$, 
\begin{equation}
\rho=\frac{\phi_2}{2v\phi}\sqrt{M^4+\left(k_1^2-k_2^2\right)^2+2M^2\left(k_1^2+k_2^2\right)},
\label{rho-arg}
\end{equation}
and one gets
\begin{equation}
\label{lambdai}
\lambda_1=\frac{\sinh \rho}{\rho}, \
\lambda_2=-\frac{\cosh \rho}{ \rho^2} 
+ \frac{\sinh \rho}{ \rho^3}, \
\lambda_3= \frac{\sinh\rho}{\rho^3} 
- 3 \frac{\cosh \rho}{\rho^4} +3 \frac{\sinh \rho}{\rho^5}.
\end{equation}
Unlike the exponent with the function $\phi_1$ in Eq.~\eqref{formfactor_phases} these factors demonstrate manifestly nonperturbative impact of the long range confining vacuum gluon fields  on  the behaviour of  form factors at short distances,  as  meson mass $M$ turns out to be entangled with with highly virtual momenta  of photons. 

An important observation is that behaviour of the functions $\lambda_i$ at large $Q^2$ is very different for symmetric ($k_1^2=k_2=Q^2$) and asymmetric ($k_1^2=Q^2$, $k_2^2=0$) kinematics since
\begin{eqnarray*}
\rho_{\rm sym}=\frac{\phi_2}{2v\phi}\sqrt{M^4+4M^2Q^2}\to \frac{\phi_2}{v\phi}M|Q|, \ \ {\rm for} \  k_1^2=k_2^2=Q^2 \gg M^2,
\\
\rho_{\rm asym}=\frac{\phi_2}{2v\phi}\ \left(M^2+Q^2\right)\to \frac{\phi_2}{2v\phi}Q^2, \ \ {\rm for} \  k_2^2=0, \ k_1^2=Q^2 \gg M^2,
\end{eqnarray*}
which results in the linear exponential increase of $\lambda_i$ as a function of $|Q|$ in the integrand of the expression for form factor in the symmetric kinematics, while quadratically growing  exponents in $\lambda_i$ are characteristic for the asymmetric one. Nevertheless, as it is shown in Appendix~\ref{appendix_asymptotic_behavior},  both regimes  demonstrate $1/Q^2$ behavior for asymptotically large $Q^2$.  However,  qualitatively different form of functions $\lambda_i$ for these kinematic regimes  results in  rather different degree of correspondence to  factorization bound 
for the coefficients in front of $1/Q^2$.

\begin{figure}
\includegraphics[width=0.49\textwidth]{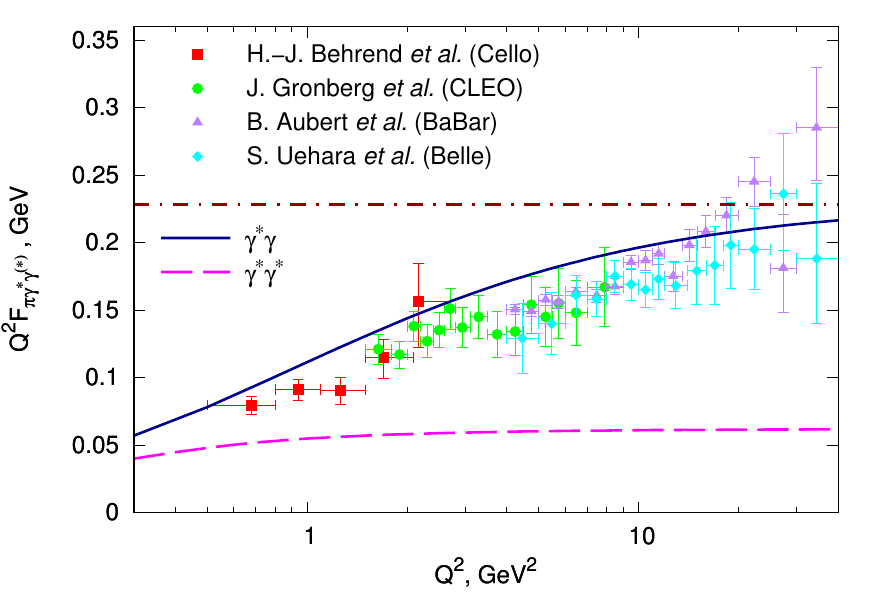}
\includegraphics[width=0.49\textwidth]{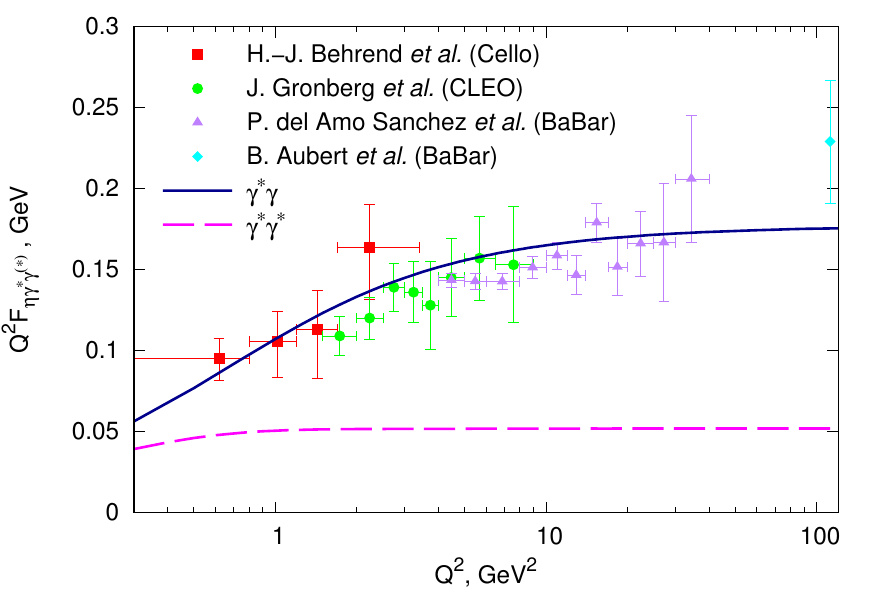}
\caption{Transition form factors of $\pi$ and $\eta$ mesons in asymmetric (solid line) and symmetric (dashed line) kinematics. The experimental data are taken from \cite{Behrend:1990sr,Gronberg:1997fj,Aubert:2009mc,Uehara:2012ag,BABAR:2011ad}.
\label{pi_eta_transition_figure}}
\end{figure}

\begin{figure}
\includegraphics[width=0.49\textwidth]{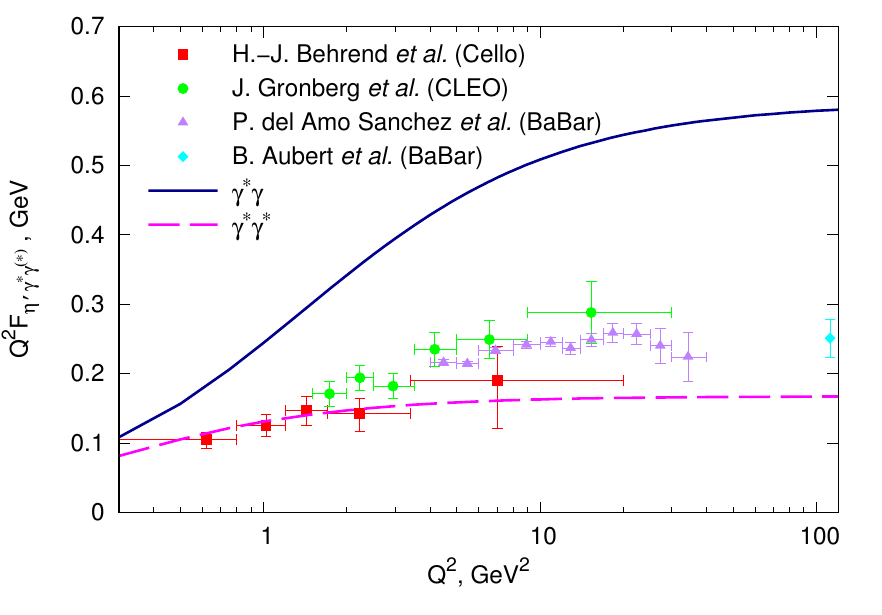}\hfill
\includegraphics[width=0.49\textwidth]{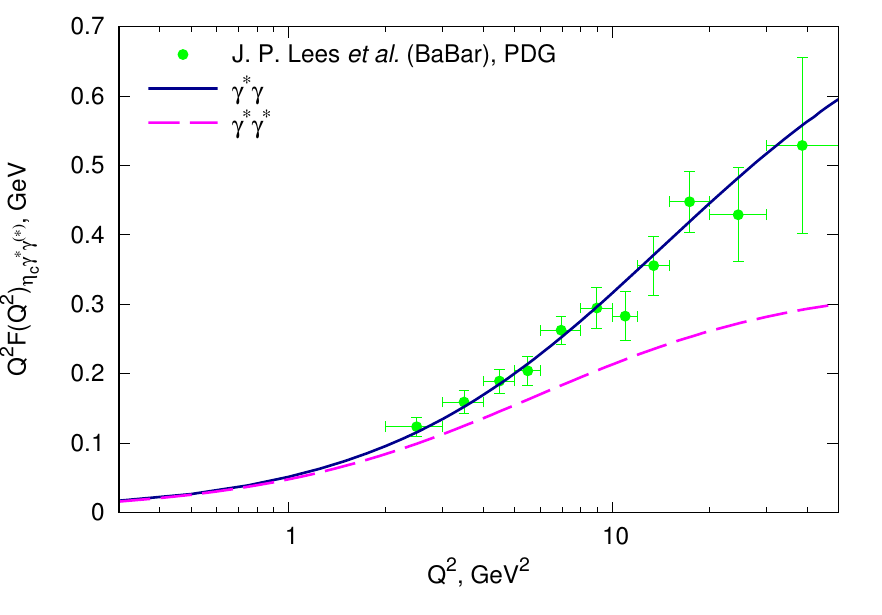}
\caption{Transition form factor for $\eta'$ (lhs) and $\eta_c$ (rhs) mesons in asymmetric kinematics (solid line) and symmetric kinematics (dashed line). Experimental data are taken from papers \cite{Behrend:1990sr,Gronberg:1997fj,BABAR:2011ad,Aubert:2009mc}
\label{eta'_eta_c_transition_figure}}
\end{figure}

Below we use notation $P$ for a couple of indices $(an)$. Transition form factor of  meson $P$ is defined as 
\begin{equation*}
F_{P\gamma^*\gamma}\left(Q^2\right)=T_P\left(-M_P^2,Q^2,0\right),
\end{equation*}
and corresponds to asymmetric kinematics with one on-shell photon. The form factor in symmetric kinematics 
\begin{equation}
\label{FP**}
F_{P\gamma^*\gamma^*}\left(Q^2\right)=T_P\left(-M_P^2,Q^2,Q^2\right),
\end{equation}
corresponds to two virtual photons.
Decay width into a couple of real photons is expressed via decay constant $g_{P\gamma\gamma}$
\begin{equation}
\Gamma(P\to \gamma\gamma)=\frac{\pi}{4}\alpha^2 M_P^3 g_{P\gamma\gamma}^2, \ 
g_{P\gamma\gamma}=T_P(-M_P^2,0,0).
\end{equation}
Experimental and calculated values of $g_{P\gamma\gamma}$ are given in Table~\ref{p2gamma_constants}.  Results of numerical computations of transition form factors for $\pi$,   $\eta$, $\eta'$ and $\eta_c$ mesons in comparison with experimental data are shown in Figs.~\ref{pi_eta_transition_figure} and \ref{eta'_eta_c_transition_figure}. It should be stressed again that all calculations   have been performed consistently with calculation of the masses, weak decay and transition constants (Fig.~\ref{fig-masses} and Table~\ref{constants}). 

Two-photon decay constants of $\pi$ and $\eta$ mesons are obtained with rather high accuracy, and overall behaviour of their form factors is reproduced quite well.  Moreover,  high $Q^2$ asymptotics   of pion form factor looks more consistent with  Belle data.  
Figure~\ref{pi_eta_transition_figure} manifestly demonstrates a qualitative difference  between the asymmetric (solid line) and symmetric (dashed line) regimes of kinematics.
 
The   constant value of $Q^2F_{\pi\gamma^*\gamma}\left(Q^2\right)$  is achieved at very large $Q^2$.  It  deviates from the factorization limit by more than 20\% (see Eq.~\eqref{pigammagamma*}). At the same time,  form factor  $Q^2F_{\pi\gamma^*\gamma^*}\left(Q^2\right)$ approaches the asymptotic value at  quite low momenta and fits factorization limit identically, as it has already been indicated  in Eq.~\eqref{pigamma*gamma*}.  Appendix~\ref{appendix_asymptotic_behavior} contains more detailed consideration of the asymptotic behaviour of the form factor. It is explicitly demonstrated that the  translation noninvariant factor~\eqref{phases}   manifestly contributes to the constant limit of $Q^2F_{\pi\gamma^*\gamma}\left(Q^2\right)$ for $Q^2\to\infty$.  This factor  is completely irrelevant for asymptotic value of $Q^2F_{\pi\gamma^*\gamma^*}\left(Q^2\right)$, which  exactly reproduces the factorization bound for symmetric kinematics.

\begin{figure}
\includegraphics[width=0.49\textwidth]{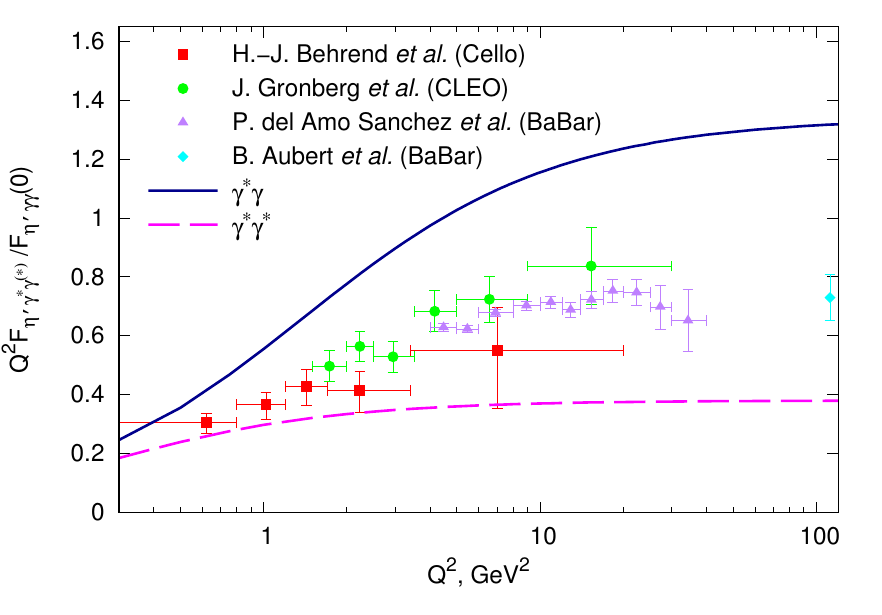}\hfill
\includegraphics[width=0.49\textwidth]{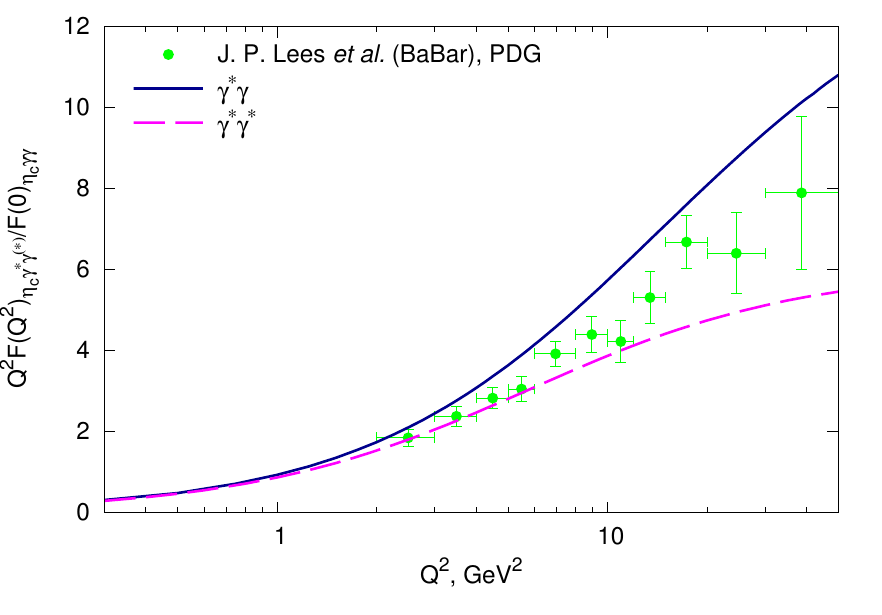}
\caption{Transition form factors for $\eta^\prime$ and $\eta_c$ mesons normalized at $Q^2=0$ and multiplied by $Q^2$. The experimental data are taken from  \cite{Behrend:1990sr,Gronberg:1997fj,BABAR:2011ad,Aubert:2009mc} and \cite{Lees:2010de} respectively. Such normalisation allows one to present and compare the overall shapes of  calculated and experimentally measured form factors in  a clear way. Mismatch of decay constants is compensated in this representattion.  \label{eta'_eta_c_transition_norm_figure}}
\end{figure}

 At this step  we  conclude that the long range confining gluon fields do not change  the conventional character of the  momentum dependence of the form factor but, depending on kinematic regime, may  influence the coefficient in front of  the  $1/Q^2$ asymptotics.  The coefficient   coincides identically with factorization prediction in the case of symmetric kinematics  in line with the analysis of papers~\cite{Dorokhov:2013xpa,Dorokhov:2010bz,Dorokhov:2010zzb}.  
Contribution of the confining Abelian (anti-)self-dual fields increases asymptotics of  $Q^2F_{\pi\gamma^*\gamma}(Q^2)$ by approximately 23 percent in comparison with the factorization limit.  This increase does not lead to contradiction with the current experimental data.  If we neglect  the main contribution of these long-range fields to  $F_{\pi\gamma^*\gamma}$ by taking $\rho=0$ in the integrand of  \eqref{formfactor_phases} than the  asymptotics  complies with the factorization bound as well (see appendix~\ref{appendix_asymptotic_behavior}). Both two-photon decay constant and the transition form factor of $\eta$ meson turn out to be also described rather satisfactory. These findings together with  previously obtained results for  masses, decay and transition constants of various mesons and new results on strong decays (see next section)  take shape of a  self-consistent picture.     
However,   consideration of  $\eta'$  and $\eta_c$  mesons  indicates that  used  approximation scheme for the quark propagators in the domain wall network background  has to be refined on explicit taking into account the inhomogeneity  of the background field at domain boundaries. 
Without implementation of this technically complicated enhancement  the picture  appears to be incomplete.  We shall discuss this difficult issue in the last section of the paper.

\section{Strong decays of  vector mesons\label{section_vpp_decays}}

Analysis of the previous section emphasized the important role  of long range confining gluon fields and related local gauge invariance of the meson effective action in formation of the transition form factors.  Strong decays of  vector mesons  allow one to verify the validity of such an  emphasis.

\begin{figure}
\begin{center}
\includegraphics[scale=0.4]{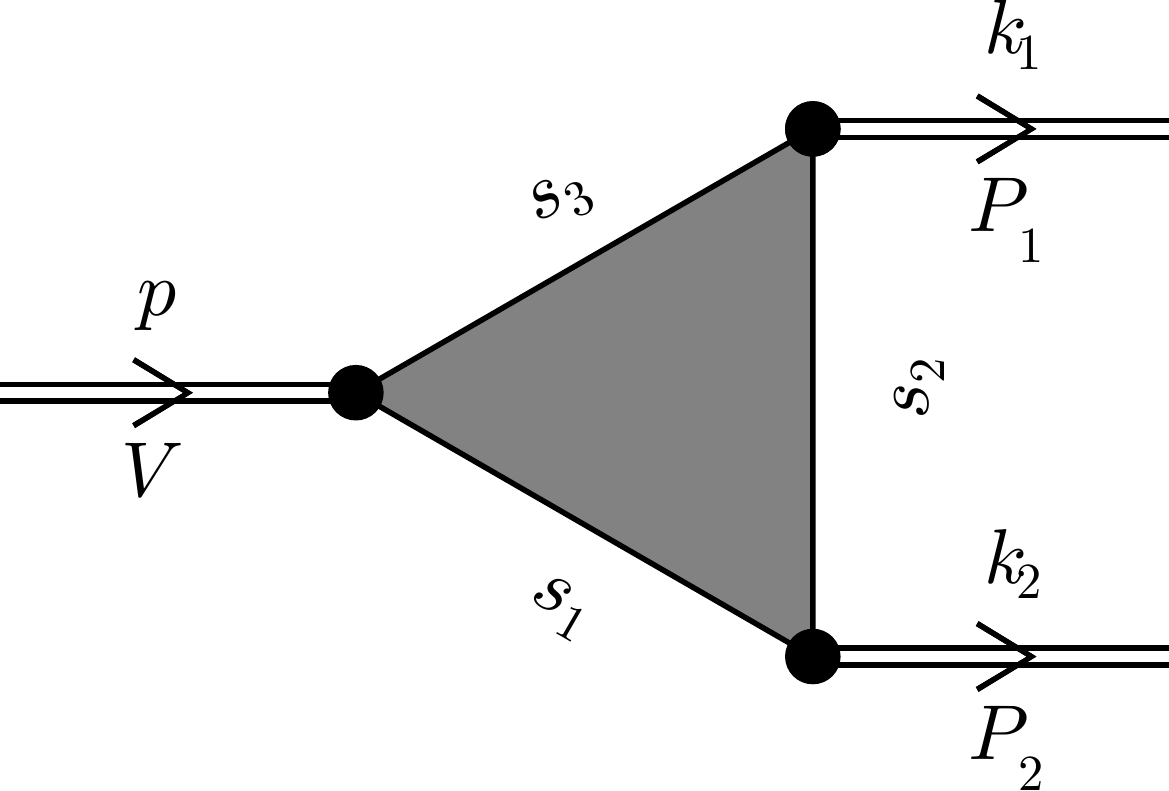}
\caption{Diagram for the three-point  meson interaction vertex. Gray background denotes averaging over background vacuum field.} \label{VPP_decay_diagram}
\end{center}
\end{figure}

The amplitude of vector meson decay into a couple of pseudoscalar mesons includes two form factors,
\begin{equation*}
A_\mu=2q_\mu A_1(p^2,q^2,pq)+p_\mu A_2(p^2,q^2,pq),
\end{equation*}
where $p_\mu$ is momentum of decaying meson, $q_\mu=k_{1\mu}-k_{2\mu}$~-
relative momentum of pseudoscalar mesons.
The corresponding decay constant is defined as the on-shell form factor
\begin{equation*}
g_{VPP}=A_1(p^2,q^2,pq), \ p^2=-M_1^2, \ k_1^2=-M_2^2, \ k_1^2=-M_3^2.
\end{equation*}
The diagram for the amplitude $A_\mu$ shown in Fig.~\ref{VPP_decay_diagram} for ground state mesons corresponds to the expressions 
\begin{gather*}
A^\mu_{aV,bP,cP}(p,k_1,k_2)=h_{aV00} h_{bP00} h_{cP00} \sum_{n_1 n_2 n_3}
\mathcal{O}^a_{0n_1}(p) \mathcal{O}^{b}_{0n_2}(k_1)
\mathcal{O}^{c}_{0n_3}(k_2) \tilde\Gamma^\mu_{aVn_1, bPn_2, cPn_3}(p,k_1,k_2),\\
\Gamma^\mu_{aVn_1, bPn_2, cPn_3}(x,y,z) = \int d\sigma_B \textrm{Tr}\gamma^{\mu}\mathcal{M}^a
F_{n_1}(x)S(x,y)i\gamma_5\mathcal{M}^b F_{n_2}(y)S(y,z)i\gamma_5\mathcal{M}^c F_{n_3}(z)S(z,x),
\end{gather*}
where we have written flavour multiplet indices $(a,b,c)$ explicitly.  In this case matrices $\mathcal{O}$ are diagonal with respect to multiplet indices, which is denoted by a single index.  The final result for the decay constant has the following
general structure
\begin{multline*}
g_{VP_1P_2}=h_Vh_{P_1}h_{P_2}
\int\limits_0^1\!\!\!\int\limits_0^1\!\!\!\int\limits_0^1\!\!\!\int\limits_0^1\!\!\!\int\limits_0^1\!\!\!\int\limits_0^1
ds_1\ ds_2\ ds_3\ dt_1\ dt_2\ dt_3 
\left(\frac{1-s_1}{1+s_1}\right)^\frac{m_1^2}{4v}
\left(\frac{1-s_2}{1+s_2}\right)^\frac{m_2^2}{4v}\left(\frac{1-s_3}{1+s_3}\right)^\frac{m_3^2}{4v}\\
\times \exp\left(\sum_{i=1}^3 \phi_i M_i^2\right) \left[ \lambda_1
+ \lambda_1\sum_{i=1}^3 F_{1i} M_i^2+\lambda_2\left( \sum_{i=1}^3
F_{2i} M_i^2 +\sum_{\substack{i,j=1,2,3\\i\leq j}}
F_{3ij} M_i^2
M_j^2\right)+\lambda_3\sum_{\substack{i,j,k=1,2,3\\i\leq j\leq
k}} F_{4ijk}M_i^2 M_j^2 M_k^2\right],
\end{multline*}
where $M_1$ is mass of vector meson, $M_2$ and $M_3$ are masses of
pseudoscalar  mesons,  $\phi$ and  $F$ are rational functions of integration variables
$s_i$ and $t_i$  independent of the meson masses.   Quark masses $m_f$ correspond to the flavour  content of the mesons. 

Functions $\lambda_i(\zeta)$ are given by the same Eq.~\eqref{f_i} as in the case of two-photon decay of pseudoscalar mesons  but now with the argument  
\begin{equation*}
\zeta=\psi\sqrt{\left(M_1^2-M_2^2-M_3^2\right)^2-4M_2^2M_3^2},
\end{equation*}
where $\psi$ is a function of the integration variables, which obviously differs from corresponding function in Eq.~\eqref{rho-arg}. 

The origin and physical content of the $\lambda$-functions discussed above in the context of the transition form factors is completely applicable to the strong $V\to PP$ decays as well. 
An important observation is that these functions turn out to be of crucial importance for  
an adequate description of the experimental results for $V\to PP$ decay constants. The results are shown in
 Table~\ref{table_vpp_decays}. Value of $g_{\omega\pi\pi}$ is exactly zero in our calculation  because of ideal mixing of $\omega$ and $\phi$ mesons and employed approximation of $SU(2)$ isospin symmetry ($m_u=m_d$). 
If one neglects translation noninvariant phases in quark propagator and drops the background field dependence in the quark-meson vertices, that means turning to just global gauge invariance of meson-meson interaction vertices,  then the argument of $\lambda_i$ has to vanish $\xi\to 0$,  and the results of calculation change dramatically as it  is seen from comparison of the third and fourth columns in Table~\ref{table_vpp_decays}.  The calculated decay constants become strongly  underestimated for light mesons  and overestimated for heavy-light mesons. This is quite known issue - it is difficult to get consistent description of both meson masses and strong decay constants.  Usually  $g_{\rho\pi\pi}$ turns out to be strongly underestimated~\cite{Bernard:1993wf,Deng:2013uca}.

The main observation of this section is that the local gauge invariance of the meson effective action is highly important for simultaneous description of meson masses and the  strong decay constants $g_{VPP}$.  This conclusion supports the analysis of the impact of the confining long range gluon fields on $P\to\gamma\gamma$ transition form factors.

\begin{table}
\begin{center}
\caption{The strong decay $V\to PP$ constants  for various decays. Here $g_{VPP}$ - result of the full calculation  with locally gauge invariant meson-meson amplitudes, $g^*_{VPP}$ -  simplified, only  globally gauge  invariant calculation.}
\begin{tabular}{c|c|c|c}
\toprule
Decay & $g^{\rm exp}_{VPP}$ \cite{PDG} & $g_{VPP}$&$g^*_{VPP}$\\
\colrule
$\rho^0\rightarrow \pi^+ \pi^-$ & $5.95$ & $7.61$&$1.14$\\
\colrule
$\omega\rightarrow \pi^+ \pi^-$ & $0.17$ & $0$&$0$\\
\colrule
$K^{*\pm} \rightarrow K^\pm \pi^0$ & $3.23$ &$3.56$&$0.65$\\
\colrule
$K^{*\pm} \rightarrow K^0 \pi^\pm$ & $4.57$ &$5.03$&$0.91$\\
\colrule
$\varphi\rightarrow K^+ K^-$ & $4.47$ &$5.69$&$1.11$\\
\colrule
$D^{*\pm}\rightarrow D^0 \pi^\pm$ & $8.41$ &$7.94$&$16.31$\\
\colrule
$D^{*\pm}\rightarrow D^\pm \pi^0$ & $5.66$ &$5.62$&$11.53$\\
\botrule
\end{tabular} \label{table_vpp_decays}
\end{center}
\end{table}

\section{Discussion and outlook
 \label{section_conclusions}}

Two-photon decay constants for $\eta_c$  and, particularly, $\eta'$ mesons  turn out to deviate from experimental values by 18 and 24 percent respectively,  which demonstrates  much less accuracy of description  than in the case of  $\pi$ and $\eta$ mesons.  
At a glance the  $\eta_c$ form factor in Fig.~\ref{eta'_eta_c_transition_figure}, calculated by means of Eq.~\eqref{formfactor_phases},  fits experimental data very well.  However, the normalized at zero momentum form factor shown in Fig.~\ref{eta'_eta_c_transition_norm_figure}  exposes its functional form  more clearly, and, though an agreement is still quite reasonable, there is a visible  excess  of  calculated form factor with respect to the experimental points.  Bearing in mind 18 percent underestimation of $g_{\eta_c\gamma\gamma}$ we  conclude that  good description seen in Fig.~\ref{eta'_eta_c_transition_figure}  is a result of mutual compensation of two inaccuracies.
 Equation \eqref{formfactor_phases} leads to strong overestimation of the normalized form factor of $\eta'$ meson as  the LHS plot in Fig.\ref{eta'_eta_c_transition_norm_figure} demonstrates. Simultaneously decay constant $g_{\eta'\gamma\gamma}$  exceeds the experimental value.
The LHS plot in Fig.~\ref{eta'_eta_c_transition_figure} demonstrates the heavy overall disagreement of $Q^2F_{\eta'\gamma^*\gamma}(Q^2)$ with experimental data.

Bearing in mind systematically good accuracy of the present approach in description of diverse properties of  various mesons  one  gets an impression, that the odd situation with $\eta'$   has to be attributed to some effects, that are particularly important for $\eta'$ meson.  If so than
it  can be plausible,  that the  triangle diagram A in Fig.~\ref{F_P_gamma_picture}  does not represent  all  potentially important contributions to the transition form factor, which are  specifically large in the case of  $\eta'$ meson. 

It is known that within the instanton liquid model the analogue of diagram B gives important contribution to the transition form factor $F_{P\gamma^*\gamma}$,   under certain conditions it can become as large as the contribution of the triangle diagram.  A detailed analysis  of this  direct instanton effects in the pion transition form factor was performed in paper~\cite{Kochelev:2009nz}. It was shown that if flavour $SU(2)$ symmetry is broken due to different current masses of up and down quarks then direct instanton effects are not small and may lead to a considerable  increase of the pion form factor at large $Q^2$. 
The addition to the form factor  is equal to the difference of  two terms related to the up- and down-quarks,  which together with other factors defines its  overall sign.  Obviously for unbroken  flavour $SU(2)$ symmetry this effect vanishes.   

This approach can be extended to all  pseudoscalar neutral mesons.
 Contribution of the diagram B to the form factors of the  $\eta^8$  and $\eta^0$ states is expected to be  nonzero for broken $SU(3)$ flavour symmetry. However there is important difference between octet and singlet states. Contribution to  $\eta^8$ form factor is given by the difference of terms related to the light quarks and the s-quark, while addition to $\eta^0$ form factor is given by a sum of terms and do not vanish even for the case of unbroken $SU(3)$.  This means that  the strongest effect  can be expected for the case of $\eta^0$  with a sign  being opposite to the sign of octet states.  In other words,  contribution of  diagram B may increase the form factors of octet states and  decrease the $\eta^0$ and $\eta_c$ form factor.  The same as for $\eta^0$ state argumentation  is applicable  to  $\eta_c$ meson  form factor but in  much more smeared  form due to the large $c$-quark mass. 

In the present approach  $\eta^0$ and $\eta^8$ ground and radially excited states are mixed with each other to form the physical $\eta$, $\eta'$ mesons and their physical radial excitations. Mixing with  $\eta_c$ is negligible. Physical meson fields diagonalize the quadratic part of the effective action \eqref{effective_meson_action}. Table~\ref{F_P_gamma_O} demonstrates the on-shell coefficients of the mixing. It should be noted that on-shell singlet-octet mixing can not be described by  one or two mixing angles since the presence of radially excited modes requires much more complicated parametrization expressed in terms of the transformation $\mathcal{O}$  (for details see~\cite{Nedelko:2016gdk}). Here it is important to note that  both diagrams shown in Fig.~\ref{diagrams} contribute to the mixing. The RHS diagram includes two-point correlator of the background gluon fields that is particularly crucial for description of the $\eta$ and $\eta'$ masses.  The diagram B in Fig.~\ref{F_P_gamma_picture} is akin to the RHS diagram in Fig.~\ref{diagrams}. The simplified scheme for taking into account the finite size of domains  in terms of  the  background field correlators  has been sufficient for estimation of $\eta$ and $\eta' $ masses, but it appears to be unable to catch the contribution to the form factor potentially arising from the  diagram B.  
Being locally (color) gauge invariant the quark polarization two-point subdiagram is  translation invariant despite the presence of translation noninvariant phases in the quark propagators. As a result of this, the  tensor structure  \eqref{tmunu},  generated initially by the presence of the (anti-)self-dual background gluon field,  turns out to be eliminated by the energy-momentum conservation in the polarization loop separately from the whole diagram.  

Thus we conclude, that the effect of direct correlations of the vacuum fields encoded in the diagam B in Fig.~\ref{F_P_gamma_picture} is  missed in our present calculation due to the  oversimplified approximation used for the quark propagators  in the presence of the domain structured background gluon field. 
Namely, the propagator in the bulk of domain with finite size is approximated by the propagator in the infinite space-time. This appears to be  an excessive simplification: the latter propagator is translation invariant up to an appropriate  gauge transformation, while the lack of translation invariance in the former one 
relates also to the finite size and random space-time position of a given domain.  
In principle this mismatch  can be improved by the use of the quark propagator represented in terms of quark  eigenmodes inside the finite domain filled by the Abelian (anti-)self-dual homogeneous gluon field analytically obtained in paper~\cite{Kalloniatis:2002ct}. At least an estimation based on the  few lowest eigenmodes (analogous to analysis of \cite{Kochelev:2009nz}) seems to be technically  realistic  task to be tackled in due course.

\section*{ACKNOWLEDGMENTS} We acknowledge fruitful discussions with
  A. Dorokhov, S. Mikhailov, N. Kochelev and O. Teryaev.

\appendix

\section{Transition form factor
\label{transition_form factor_formula}}

Polynomial functions   $F_i(s_1,s_2,s_3,t)$ in Eq.~\eqref{formfactor_phases} have the following form:

\begin{multline*}
F_1 = -2 \left(2 s_1^4 \left(2 (s_2-1) (s_2+1) t^2 v^2
    \left(s_2^2 \left(s_3^2-2\right)+s_2 s_3
    \left(s_3^2-1\right)+s_3^2 \left(2
    s_3^2-1\right)\right)+\right.\right.\\
\left.\left.   +t v \left(s_2^4 s_3
    \left(s_3^2-4\right)-2 s_2^3
    \left(s_3^4-s_3^2+1\right)+s_2^2 s_3
    \left(s_3^2+3\right)+2 s_2 s_3^2-2
    s_3^3+s_3\right)-\right.\right.\\
\left.\left. -s_2 s_3 (s_2-s_3)
    (s_2+s_3) (s_2 s_3+1)\right)+s_1^3
    \left((s_2-1) (s_2+1) t v \left(s_2^2 \left(5
    s_3^4+3 s_3^2-4\right)+\right.\right.\right.\\
\left.\left.\left. +s_2 \left(5 s_3^2-9\right)
    s_3+4 s_3^4-4 s_3^2-2\right)+2 (s_2-1)
    (s_2+1) t^2 v^2 \left(5 s_2^2
    \left(s_3^2-1\right) s_3+s_2
    \left(s_3^4+s_3^2-2\right)+6 s_3^3-4 s_3\right)\right.\right.\\
\left.\left.+2
    \left(s_2 \left(-s_2^3 s_3+2 s_2
    \left(s_2^2-1\right) s_3^3-\left(s_2^2+1\right)
    s_3^2+s_3^4+1\right)+s_3\right)\right)+s_1^2
    \left(2 s_2^4 s_3^2-2 s_2^2 \left(s_3^4+3
    s_3^2-2\right)+\right.\right.\\
\left.\left. +2 (s_2-1) (s_2+1) t^2 v^2
    \left(s_2^2 \left(3 s_3^4-s_3^2+2\right)+s_2
    \left(s_3-s_3^3\right)-2 s_3^4+2
    s_3^2-2\right)+t v \left(4
    \left(s_2^3+s_2\right)+\right.\right.\right.\\
\left.\left.\left.+s_2 \left(3 s_2^2+5\right)
    s_3^4-s_2 \left(7 s_2^2+9\right) s_3^2+\left(7
    s_2^4-11 s_2^2+4\right) s_3^3+\left(s_2^4-3
    s_2^2+2\right) s_3\right)-4 s_2 s_3
    \left(s_3^2-1\right)+4 s_3^2-2\right)\right.\\
\left.+2 s_1
    \left(s_2^4 \left(-s_3^3\right)-s_2^3
    s_3^4+s_2^3 s_3^2+s_2^3-2 (s_2-1)
    (s_2+1) t^2 v^2 \left(\left(2 s_2^2+3\right)
    s_3^3-2 \left(s_2^2+1\right) s_3+s_2
    s_3^4-s_2\right)-\right.\right.\\
\left.\left.-(s_2-1) (s_2+1) t v
    \left(s_2^2 \left(3 s_3^4-1\right)+s_2 \left(3
    s_3^2-5\right) s_3+s_3^4+2 s_3^2-4\right)+2
    s_2^2 s_3+s_2 s_3^2-2 s_2+s_3^3-2
    s_3\right)+\right.\\
\left.+2 t v \left(s_2^4 \left(s_3-2
    s_3^3\right)+2 s_2^3 \left(-s_3^4+s_3^2+1\right)+3
    s_2^2 s_3+2 s_2 \left(s_3^2-2\right)+2 s_3
    \left(s_3^2-2\right)\right)-\right.\\
\left.-4 (s_2-1) (s_2+1) t^2
    v^2 \left(s_3 \left(s_2 \left(s_2 \left(2
    s_3^2-1\right) s_3+s_3^2-1\right)+2
    s_3\right)-2\right)+2 (s_2+s_3)^2 (s_2
    s_3-1)\right)
\end{multline*}
\begin{multline*}
F_2 = -2 \left(2 \left(\left(2 \left(s_3^2-3\right) s_3^2-\left(7
    s_3^2+4\right) t v s_3+2 \left(2-5 s_3^2\right)
    t^2 v^2\right) s_2^4+\left(s_3 \left(4 s_3^2+6
    \left(s_3^2-1\right) t v s_3\right.\right.\right.\right.\\
\left.\left.\left.\left.+6
    \left(s_3^2-1\right) t^2 v^2-5\right)-2 t v\right)
    s_2^3+\left(4 s_3^4+2 s_3^2+\left(13 s_3^2+7\right)
    t v s_3+4 \left(s_3^4+2\right) t^2 v^2+1\right)
    s_2^2+\right.\right.\\
\left.\left.+\left(s_3^3-14 \left(s_3^2-1\right) t^2 v^2
    s_3+2 \left(-2 s_3^4+s_3^2+2\right) t v\right)
    s_2-3 s_3 (s_3+t v) (2 s_3 t
    v+1)\right) s_1^4+\right.\\
\left.+\left(2 s_3 \left(6 s_3^2-5\right)
    s_2^4+2 \left(4 s_3^4-5 s_3^2+2\right)
    s_2^3+\left(8 s_3^3-4 s_3\right) s_2^2+2
    \left(s_3^4-s_3^2-1\right) s_2-12 s_3^3+\right.\right.\\
\left.\left.+2 \left(-17
    s_3 \left(s_3^2-1\right) s_2^4+3
    \left(s_3^4+s_3^2-2\right) s_2^3+s_3 \left(11-5
    s_3^2\right) s_2^2+\left(-11 s_3^4-3
    s_3^2+14\right) s_2+\right.\right.\right.\\
\left.\left.\left.+6 s_3
    \left(s_3^2-2\right)\right) t^2 v^2+6 s_3-(s_2-1)
    (s_2+1) \left(-12 s_3^4+s_2 \left(27-31
    s_3^2\right) s_3+s_2^2 \left(5 s_3^4-13
    s_3^2+4\right)-6\right) t v\right) s_1^3+\right.\\
\left.+\left(\left(-6
    s_3^4+10 s_3^2+\left(25 s_3^2-1\right) t v
    s_3-2 \left(7 s_3^4-29 s_3^2+10\right) t^2
    v^2+2\right) s_2^4-\right.\right.\\
\left.\left.-\left(s_3^2-1\right) (4 t
    v+s_3 (3 t v (5 s_3+2 t v)+4))
    s_2^3-\left(s_3 \left(17 t v+s_3 \left(2
    s_3^2+25 t v s_3+6 \left(s_3^2+5\right)
    t^2 v^2+6\right)\right)+4\right)
    s_2^2+\right.\right.\\
\left.\left.+\left(s_3^2-1\right) (12 t v+s_3 (t
    v (7 s_3+6 t v)-8)) s_2-12 \left(s_3^4-3
    s_3^2+1\right) t^2 v^2-6 s_3^2
    \left(s_3^2-2\right)+18 s_3 t v\right) s_1^2+\right.\\
\left.+2
    \left(-5 s_3^3 s_2^4+4 s_3 s_2^4-5 s_3^4
    s_2^3+5 s_3^2 s_2^3-s_2^3-2 s_3^3
    s_2^2+s_3^2 s_2+3 s_3^3+2 \left(10 s_3
    \left(s_3^2-1\right) s_2^4-3 \left(s_3^4-1\right)
    s_2^3+\right.\right.\right.\\
\left.\left.\left.+s_3 \left(s_3^2-4\right) s_2^2+7
    \left(s_3^4-1\right) s_2-3 s_3^3+6 s_3\right)
    t^2 v^2-(s_2-1) (s_2+1)
    \left(\left(s_2^2-3\right) s_3^4+9 s_2 s_3^3+2
    \left(s_2^2+6\right) s_3^2-\right.\right.\right.\\
\left.\left.\left.-7 s_2
    s_3-s_2^2\right) t v\right) s_1+2 s_2
    \left(s_3 (s_2 s_3-1) (s_2+s_3)^2+2 \left(2
    s_2 \left(s_2^2+2\right) s_3^4+\left(7-3
    s_2^2\right) s_3^3-s_2 \left(5 s_2^2+3\right)
    s_3^2+\right.\right.\right.\\
\left.\left.\left.+\left(3 s_2^2-7\right) s_3+2 s_2\right)
    t^2 v^2+\left(4 s_3^4-6 s_3^2+s_2 \left(2
    s_3^2-1\right) s_3+s_2^3 \left(s_3-2
    s_3^3\right)+2 s_2^2
    \left(-s_3^4+s_3^2+1\right)\right) t v\right)\right)
\end{multline*}
\begin{equation*}
F_3 = \frac{4}{v} (s_1 (s_2-s_3)+s_2 s_3-1)
    (s_1 s_2 s_3+s_1+s_2+s_3) (t v(s_1+s_3)+s_1 s_3)s_2
\end{equation*}
\begin{multline*}
F_4 = -\frac{1}{v}2s_2 (s_1 (s_2-s_3)+s_2 s_3-1)
    (t v (s_1+s_3)+s_1 s_3) \left(s_1^3
    s_2^3 s_3^3+s_1^3 s_2^2 s_3^2-2 s_1^3
    s_2 s_3^3+s_1^3 s_2 s_3-2 s_1^3
    s_3^2+s_1^3+\right.\\
\left.+s_1^2 s_2^3 s_3^2+s_1^2
    s_2^2 s_3^3+2 t v (s_1 s_2
    s_3+s_1+s_2+s_3) \left(s_1^2 (s_2
    (s_3 (s_2-2 s_3)+2)-s_3)+\right.\right.\\
\left.\left.+s_1
    \left(s_2^2-1\right) \left(s_3^2-1\right)-s_2 (s_3
    (s_2-2 s_3)+2)+s_3\right)+2 \left(s_1^2-1\right)
    \left(s_3^2-1\right) t^2 v^2 \left(s_2 s_3
    \left(s_1 \left(s_2^2-3\right)-3 s_2\right)-\right.\right.\\
\left.\left.-3 s_1
    s_2^2+s_1+s_2^3-3 s_2+s_3\right)-2
    s_1^2 s_2 s_3^2+s_1^2 s_2-2 s_1^2
    s_3^3+s_1^2 s_3-s_1 s_2^3 s_3-s_1
    s_2^2+\right.\\
\left.+s_1 s_2 s_3^3+s_1
    s_3^2-s_2^3-s_2^2 s_3+s_2
    s_3^2+s_3^3\right)
\end{multline*}
\begin{multline*}
F_5 = -\frac{1}{v} 2s_2 (s_1 (s_2-s_3)+s_2 s_3-1)
    (t v (s_1+s_3)+s_1 s_3) \left(-s_1^3
    s_2^3 s_3^3+2 s_1^3 s_2^3
    s_3-s_1^3 s_2^2 s_3^2+2 s_1^3
    s_2^2-s_1^3 s_2 s_3-\right.\\
\left.-s_1^3-s_1^2
    s_2^3 s_3^2+2 s_1^2 s_2^3-s_1^2 s_2^2
    s_3^3+4 \left(s_1^2-1\right) s_2
    \left(s_3^2-1\right) t^2 v^2 \left(s_1
    \left(s_2^2-2\right) s_3-s_1
    s_2+s_2^2-s_2 s_3-2\right)+\right.\\
\left.+2 t v (s_1
    s_2 s_3+s_1+s_2+s_3) \left(s_1^2
    (s_2 (s_3 (s_2-2 s_3)+2)-s_3)-s_1
    \left(s_2^2-1\right) \left(s_3^2-1\right)-\right.\right.\\
\left.\left.-s_2 (s_3
    (s_2-2 s_3)+2)+s_3\right)+2 s_1^2 s_2^2
    s_3-s_1^2 s_2-s_1^2 s_3-s_1 s_2^3
    s_3-s_1 s_2^2+s_1 s_2 s_3^3+\right.\\
\left.+s_1
    s_3^2-s_2^3-s_2^2 s_3+s_2
    s_3^2+s_3^3\right)
\end{multline*}
\begin{equation*}
F_6(s_1,s_2,s_3,t)=F_5(s_3,s_2,s_1,t)
\end{equation*}
\begin{multline*}
F_7 = -\frac{1}{2 v^2}s_2^2 (t v (s_1+s_3)+s_1 s_3)^2
    \left(2 s_1^4 \left(s_2^4 s_3 (s_3 (t v (3
    s_3+2 t v)+2)+2 t v)+s_2^3 s_3 \left(1-2
    t v \left(3 \left(s_3^2-1\right) t v \right.\right.\right.\right.\\
\left.\left.\left.\left.+s_3
    \left(s_3^2-2\right)\right)\right)+s_2^2 \left(-2
    s_3^4-\left(5 s_3^2+3\right) s_3 t v-4
    \left(s_3^4-s_3^2+1\right) t^2 v^2-1\right)+s_2
    \left(-s_3^3+6 \left(s_3^2-1\right) s_3 t^2 v^2-2
    t v\right)+\right.\right.\\
\left.\left.+s_3 (s_3+t v) (2 s_3 t
    v+1)\right)+s_1^3 \left(2 (s_2-1) (s_2+1) t^2 v^2
    \left(3 s_2^2 s_3 \left(s_3^2-1\right)-s_2 \left(5
    s_3^4+s_3^2-6\right)+2 s_3
    \left(s_3^2-2\right)\right)+\right.\right.\\
\left.\left.+(s_2-1) (s_2+1) t v
    \left(s_2^2 \left(3 s_3^4+s_3^2\right)+s_2
    \left(9-5 s_3^2\right) s_3-4 s_3^4-2\right)+2
    \left(s_2^4 s_3+s_2^3 \left(3 s_3^2-2\right)-2
    s_2^2 s_3^3-\right.\right.\right.\\
\left.\left.\left.-s_2 \left(s_3^4+s_3^2-1\right)+2
    s_3^3-s_3\right)\right)+s_1^2 \left(s_2^4
    \left(-\left(3 s_3^2+5\right) s_3 t v+2
    \left(s_3^2-1\right) s_3^2+2 \left(s_3^4-7
    s_3^2+2\right) t^2 v^2-2\right)+\right.\right.\\
\left.\left.+s_2^3 s_3
    \left(s_3^2-1\right) (t v (9 s_3+2 t
    v)+4)+s_2^2 \left(\left(3 s_3^2+11\right) s_3 t
    v+2 \left(s_3^2-1\right) s_3^2+2 \left(5 s_3^4-3
    s_3^2+4\right) t^2 v^2+4\right)-\right.\right.\\
\left.\left.-s_2
    \left(s_3^2-1\right) t v \left(s_3^2+2 s_3
    t v+8\right)+2 s_3^2 \left(s_3^2-2\right)+4
    \left(s_3^4-3 s_3^2+1\right) t^2 v^2-6 s_3
    t v\right)+\right.\\
\left.+2 s_1 \left(s_2^4 s_3^3-2 s_2^4
    s_3+s_2^3 s_3^4-3 s_2^3
    s_3^2+s_2^3+(s_2-1) (s_2+1) t v
    \left(s_2^2 \left(s_3^4-2 s_3^2-1\right)+s_2
    \left(3 s_3^2-5\right) s_3-s_3^4+4 s_3^2\right)-\right.\right.\\
\left.\left.-2
    (s_2-1) (s_2+1) t^2 v^2 \left(2 s_2^2
    \left(s_3^2-1\right) s_3-3 s_2
    \left(s_3^4-1\right)+s_3^3-2 s_3\right)+2 s_2^2
    s_3+s_2 s_3^2-s_3^3\right)-\right.\\
\left.-2 s_2 t v
    \left(s_2 \left(s_2^2-1\right) s_3+2 s_2^2+2
    s_3^4-4 s_3^2\right)+4 s_2 t^2 v^2 \left(3
    \left(s_2^2-1\right) s_3^3+s_2 \left(s_2^2+3\right)
    s_3^2-3 \left(s_2^2-1\right) s_3-\right.\right.\\
\left.\left.-2 s_2
    s_3^4-2 s_2\right)-2 s_2 s_3
    (s_2+s_3)^2 (s_2 s_3-1)\right).
\end{multline*}

\section{Asymptotic behavior of the form factor
\label{appendix_asymptotic_behavior}}

In essence, calculation of the asymptotic behaviour of the form factor \eqref{formfactor_phases} at $p^2=-M^2$  is reduced to consideration of the  integral 
\begin{eqnarray}
\label{form factor_phases}
I^{fn}(M^2,k_1^2,k_2^2)&=&\int_0^1 ds_1 \int_0^1 ds_2 
\int_0^1 ds_3 \int_0^1 dt\ t^{n} \frac{\partial^{n}}{\partial t^{n}}\exp \left\{-\frac{k_1^2 \phi_{11}+k_2^2\phi_{12}}{2v\phi}+\frac{M^2\phi_1}{4v\phi}\right\}\\
&\times &\left[\left(\frac{1-s_1}{1+s_1}\right)\left(\frac{1-s_2}{1+s_2}\right) 
\left(\frac{1-s_3}{1+s_3}\right)\right]^{m_f^2/4v} 
\frac{1}{(1-s_1^2)(1-s_2^2)(1-s_3^2)}
\nonumber\\
&\times & \frac{1}{\phi^2}
\left[ \lambda_1 \frac{F_1}{\phi^2}+\lambda_2 
\left(\frac{F_2}{\phi^2}+m_f^2\frac{F_3}{\phi} +M^2\frac{F_4}{\phi^3} 
+k_1^2\frac{F_5}{\phi^3}+k_2^2\frac{F_6}{\phi^3}\right)+\lambda_3 
\left[\left(M^2+k_1^2+k_2^2\right)^2-4k_1^2k_2^2\right]\frac{ 
F_7}{\phi^4} \right]
\nonumber
\end{eqnarray}
in different regimes over momenta $k_1^2$ and $k_2^2$.  Here and below dimensionless notations for momenta and masses like $M\equiv M/\Lambda$ are used. It is convenient to use the following representation for $\lambda_i$-functions  \eqref{lambdai}
\begin{gather*}
\lambda_1(\rho)=\frac{1}{2}\int_{-1}^1 d\kappa \exp{\kappa \rho},\quad
\lambda_2(\rho)=\frac{1}{4} \int_{-1}^1 d\kappa \left(\kappa^2 -1\right)\exp\kappa \rho,\quad
\lambda_3(\rho)=\frac{1}{16} \int_{-1}^1 d\kappa (\kappa^2-1)^2 \exp \kappa \rho,
\end{gather*}
in the integral~\eqref{form factor_phases} in order to  represent it as
\begin{eqnarray}
\nonumber
I^{fn}(M^2,k_1^2,k_2^2)&=&\int_{-1}^1 d\kappa \int_0^1 ds_1 \int_0^1 ds_2 
\int_0^1 ds_3 \int_0^1 dt\ t^{n} \frac{\partial^{n}}{\partial t^{n}}\\
&\times& \exp \left\{-\frac{k_1^2 \phi_{11}+k_2^2\phi_{12}}{2v\phi}+\frac{M^2\phi_1}{4v\phi}+\kappa \frac{\phi_2}{2v\phi}\sqrt{M^4+\left(k_1^2-k_2^2\right)^2+2M^2\left(k_1^2+k_2^2\right)}\right\}\\
&\times &\left[\left(\frac{1-s_1}{1+s_1}\right)\left(\frac{1-s_2}{1+s_2}\right) 
\left(\frac{1-s_3}{1+s_3}\right)\right]^{m_f^2/4v} 
\frac{1}{(1-s_1^2)(1-s_2^2)(1-s_3^2)}
\nonumber\\
&\times &\frac{1}{\phi^2}
\left[  \frac{F_1}{2\phi^2}+\frac{1}{4}(\kappa^2-1)
\left(\frac{F_2}{\phi^2}+m_f^2\frac{F_3}{\phi} +M^2\frac{F_4}{\phi^3} 
+k_1^2\frac{F_5}{\phi^3}+k_2^2\frac{F_6}{\phi^3}\right)
\right.
\nonumber\\
&+& \left.\frac{1}{16}(\kappa^2-1)^2
\left[\left(M^2+k_1^2+k_2^2\right)^2-4k_1^2k_2^2\right]\frac{ 
F_7}{\phi^4} \right].
\label{form factor_phases-1}
\end{eqnarray}
Below  two  different kinematic regimes are considered.

\subsection{Asymmetric kinematics: $k_1=Q^2,k_2^2=0$}
In this case the integral reads
\begin{eqnarray}
\label{form factor_phases-2}
I^{fn}(M^2,Q^2,0)&=&\int_{-1}^1 d\kappa\int_0^1 ds_1 \int_0^1 ds_2 
\int_0^1 ds_3 \int_0^1 dt\ t^{n} \frac{\partial^{n}}{\partial t^{n}}\exp \left\{-\frac{Q^2 \phi_{11}}{2v\phi}+\frac{M^2\phi_1}{4v\phi}+\kappa \frac{\phi_2}{2v\phi}\left(M^2+Q^2\right)\right\}\\
&\times &\left[\left(\frac{1-s_1}{1+s_1}\right)\left(\frac{1-s_2}{1+s_2}\right) 
\left(\frac{1-s_3}{1+s_3}\right)\right]^{m_f^2/4v} 
\frac{1}{(1-s_1^2)(1-s_2^2)(1-s_3^2)}
\nonumber\\
&\times &\frac{1}{\phi^2}
\left[ \frac12 \frac{F_1}{\phi^2}+\frac{1}{4}(\kappa^2-1)
\left(\frac{F_2}{\phi^2}+m_f^2\frac{F_3}{\phi} +M^2\frac{F_4}{\phi^3} 
+Q^2\frac{F_5}{\phi^3}\right)+\frac{1}{16}(\kappa^2-1)^2
\left(M^2+Q^2\right)^2\frac{ 
F_7}{\phi^4} \right].
\nonumber
\label{phi1-1}
\end{eqnarray}

For studying the large $Q^2$ limit it is convenient to separate dependence on $s_2$  and define polynomials $\chi_i$ which depend  on other integration variables,
\begin{eqnarray*}
\chi_1=s_1+tv +s_1s_3 tv,\ 
\chi_2=s_1s_3+(s_1+s_3)tv,\\
\chi_3=2s_1s_3(1+s_2\kappa)+tv(s_1-s_2+s_3 -s_1s_2s_3 +2s_2\kappa(s_1+s_3)),
\end{eqnarray*}
so that the expression in the exponent takes the form:
\begin{eqnarray*}
-\frac{Q^2 \phi_{11}}{2v\phi}+\frac{M^2\phi_1}{4v\phi}+\kappa \frac{\phi_2}{2v\phi}\left(M^2+Q^2\right)= -\frac{ s_2\left(\chi_1-\kappa\chi_2\right)}{2v\phi}Q^2+\frac{\chi_3}{4v\phi}M^2.
\end{eqnarray*}
One can see from the argument of exponent in \eqref{form factor_phases-2} that asymptotic behavior of the integral at large $Q^2$ is determined by the vicinity of  $s_2=0$, which corresponds to the ultraviolet (short distance) regime in the translation invariant part quark propagator \eqref{quark_propagator_trinv} between two photons, see Fig.~\ref{F_P_gamma_picture}. The leading short distance form of $H_f$   is not modified by the background field, it  simply  corresponds to the standard free Euclidean quark propagator. 
However the translation noninvariant phase in full propagator \eqref{quark_propagator} may mix up the short and large distance regimes in the diagram \ref{F_P_gamma_picture} as whole thus leading to a modification of the asymptotic behaviour of the form factor at large momenta.  

Keeping only the lowest-order terms in the expansion of the integrand at  $s_2=0$   one arrives at the
asymptotics  of the integral,
\begin{eqnarray*}
I^{fn}(-M^2,Q^2,0) \sim {{\mathcal I}^{fn }_{\gamma^*\gamma}}/{Q^2},
\end{eqnarray*} 
with $Q$-independent coefficient
\begin{eqnarray}
{\mathcal I}^{fn }_{\gamma^*\gamma}=\int_0^1 ds_1 \int_0^1 ds_3 \int_{-1}^1 d\kappa \int_0^1 dt\ t^n \frac{\partial^n}{\partial t^n}
\left[\left(\frac{1-s_1}{1+s_1}\right)\left(\frac{1-s_3}{1+s_3}\right)\right]^{\frac{m_f^2}{4v}} \frac{1}{(1-s_1^2)(1-s_3^2)}
\exp \left\{M^2\frac{\chi_3}{4v\phi}\right\}
\label{nonsym}\\
\times \frac{2v}{\left(\chi_1-\kappa\chi_2\right)\phi}\left[ \frac12 \frac{F_1}{\phi^2}+\frac14\left(\kappa^2-1\right) \left(\frac{F_2}{\phi^2} +\frac{2v}{(\chi_1-\kappa\chi_2)\phi^2}\frac{\partial F_5}{\partial s_2}\right)\left.+\frac{1}{16}\left(\kappa^2-1\right)^2  \frac{4v^2 }{(\chi_1-\kappa\chi_2)^2\phi^2}\frac{\partial^2 F_7}{\partial s_2^2} \right]\right|_{s_2=0}.
\nonumber
\end{eqnarray}
The variable $\kappa$ can  be integrated out analytically:
\begin{gather*}
\int_{-1}^1 d\kappa \frac{1}{\chi_1-\kappa \chi_2}=\frac{1}{\chi_2}\ln\frac{\chi_1+\chi_2}{\chi_1-\chi_2},\\
\int_{-1}^1 d\kappa \frac{\kappa^2-1}{\chi_1-\kappa \chi_2}=-2\frac{\chi_1}{\chi_2^2}+\frac{\chi_1^2}{\chi_2^3}\ln\frac{\chi_1+\chi_2}{\chi_1-\chi_2}- \frac{1}{\chi_2} \ln\frac{\chi_1+\chi_2}{\chi_1-\chi_2},\\
\int_{-1}^1 d\kappa \frac{\kappa^2-1}{\left(\chi_1-\kappa \chi_2\right)^2}=\frac{4}{\chi_2^2} -2\frac{\chi_1}{\chi_2^3}\ln\frac{\chi_1+\chi_2}{\chi_1-\chi_2},\\
\int_{-1}^1 d\kappa \frac{\left(\kappa^2-1\right)^2}{\left(\chi_1-\kappa \chi_2\right)^3}=-12\frac{\chi_1}{\chi_2^4} +6\frac{\chi_1^2}{\chi_2^5}\ln\frac{\chi_1+\chi_2}{\chi_1-\chi_2}- \frac{2}{\chi_2^3}\ln\frac{\chi_1+\chi_2}{\chi_1-\chi_2},\\
\frac{\chi_1+\chi_2}{\chi_1-\chi_2}=\frac{1+s_3}{1-s_3}\frac{s_1+tv+s_1tv}{s_1+tv-s_1tv}.
\end{gather*}
After substitution of this result to Eq.~\eqref{form factor_phases-1}, 
the rest of integrals  can be calculated numerically. The result contributes to coefficient  $\varkappa_{\gamma^*\gamma}=1.23$ in Eq.~\eqref{pigammagamma*}.

If we now neglect the main effect of the background field  by eliminating the term $\kappa*\chi_2$
in Eq.~\eqref{nonsym}, then numerical calculation of the asymptotics gives the value $\varkappa_{\gamma^*\gamma}=1.014$  which agrees with  factorization limit very well.

\subsection{Symmetric kinematics:  $k_2^2=k_1^2=Q^2$}
The integral takes the form

\begin{eqnarray}
\label{form factor_phases-3}
I^{fn}(M^2,Q^2,Q^2)&=&\int_{-1
}^1 d\kappa\int_0^1 ds_1 \int_0^1 ds_2 
\int_0^1 ds_3 \int_0^1 dt\ t^{n} \frac{\partial^{n}}{\partial t^{n}}\\
&\times& \exp \left\{-\frac{\phi_{11}+\phi_{12}}{2v\phi}Q^2+\frac{\phi_1}{4v\phi}M^2+\kappa \frac{\phi_2}{2v\phi}\sqrt{M^4+4M^2 Q^2}\right\}
\nonumber\\
&\times &\left[\left(\frac{1-s_1}{1+s_1}\right)\left(\frac{1-s_2}{1+s_2}\right) 
\left(\frac{1-s_3}{1+s_3}\right)\right]^{m_f^2/4v} 
\frac{1}{(1-s_1^2)(1-s_2^2)(1-s_3^2)}
\nonumber\\
&\times &\frac{1}{\phi^2}
\left[ \frac12 \frac{F_1}{\phi^2}+\frac{1}{4}(\kappa^2-1)
\left(\frac{F_2}{\phi^2}+m_f^2\frac{F_3}{\phi} +M^2\frac{F_4}{\phi^3} 
+Q^2\frac{F_5+F_6}{\phi^3}\right)
\right.\nonumber\\
&+&\left.\frac{1}{16}(\kappa^2-1)^2
\left(M^4+4M^2Q^2\right)\frac{ 
F_7}{\phi^4} \right].
\nonumber
\label{phi1-2}
\end{eqnarray}
For $Q^2\gg M^2$ the second term in the exponent in Eq.~\eqref{form factor_phases-3} is subleading with respect   to $\phi_1$ (see \eqref{phi1-2}) as it is linear in $|Q|$ .   In other words, term~\eqref{phases} does not contribute to asymptotic behavior of the form factor in symmetric kinematics, that is the crucial difference between asymmetric and symmetric kinematic regimes.

\begin{figure}
\begin{center}
\includegraphics[scale=1.]{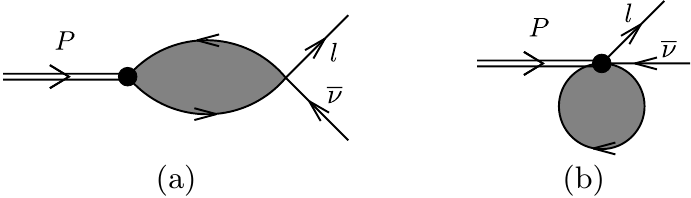}
\caption{Diagrams for the weak decay constant of charged pseudoscalar mesons.} \label{fP_weak_decay_diagram}
\end{center}
\end{figure}

\begin{eqnarray*}
{\mathcal I}^{fn }_{\gamma^*\gamma^*} = 2v\int_0^1 ds_1 \int_0^1 ds_3 \int_0^1 dt\ t^n \frac{\partial^n}{\partial t^n}\left[\left(\frac{1-s_1}{1+s_1}\right)\left(\frac{1-s_3 }{1+s_3}\right)\right]^{\frac{m_f^2}{4v}} \frac{1}{(1-s_1^2)(1-s_3^2)}
\\
\times \frac{1}{\phi^4}\left[ F_1- \frac{1}{3}F_2  \right]
 \left.\exp \left\{M^2\frac{\phi_1}{4v\phi}\right\}
\right|_{s_2=0}.
 \end{eqnarray*}
 Taking into account  the explicit form of the polynomials $F_1$, $F_2$, $\phi$ and $\phi_1$ at $s_2=0$ one arrives at
\begin{eqnarray}
{\mathcal I}^{fn }_{\gamma^*\gamma^*}=8v\int_0^1 ds_1 \int_0^1 ds_3 \int_0^1 dt\ t^n \frac{\partial^n}{\partial t^n}\left[\left(\frac{1-s_1}{1+s_1}\right)\left(\frac{1-s_3}{1+s_3}\right)\right]^{\frac{m_f^2}{4v}} \frac{1}{(1-s_1^2)(1-s_3^2)}
\nonumber\\
\times 
 \frac{ (1+s_1s_3)[(1 -s_1s_3)(s_1+s_3)+2vt(1-s_1^2)(1-s_3^2)]}{(s_1+s_3 +2vt(1+s_1 s_3) )^3} 
\nonumber \\
 \times \exp \left\{M^2\frac{2s_1s_3+vt(s_1+ s_3 )}{4v(s_1+s_3 +2vt(1+s_1 s_3))}\right\}.
 \label{I**}
\end{eqnarray}
 Substituting this expression into~\eqref{formfactor_phases}, one obtains for the form factor of the $a$-th component of the flavour  pseudoscalar multiplet  with radial excitation number $n$ 
\begin{equation}
Q^2F_{P_{an}\gamma^*\gamma^*}(Q^2)\sim
\Lambda\frac{h_{aP0n}}{16\pi^2}\sum_{n^\prime,f,v}\frac{1}{v}\mathcal{O}^{ab}_{nn^\prime}  \mathcal{M}_{ff}^b
q_f^2m_f  
{\mathcal I}^{fn' }_{\gamma^*\gamma^*}.
\end{equation}
In particular, for $\pi$-meson which corresponds to  $a=3$ and $n=0$  one arrives at 
\begin{equation}
Q^2F_{\pi\gamma^*\gamma^*}(Q^2)\sim \frac{\sqrt{2}}{3}f_\pi
\end{equation}
with the constant  
\begin{eqnarray} 
f_\pi= m  \frac{h_{\pi}}{4\pi^2}\sum_{n,v}\mathcal{O}^{33}_{0n} 
\int_0^1 ds_1 \int_0^1 ds_3 \int_0^1 dt\ t^n \frac{\partial^n}{\partial t^n}\left[\left(\frac{1-s_1}{1+s_1}\right)\left(\frac{1-s_3}{1+s_3}\right)\right]^{\frac{m^2}{4v\Lambda^2}} \frac{1}{(1-s_1^2)(1-s_3^2)}
\nonumber\\
\times 
 \frac{ (1+s_1s_3)[(1 -s_1s_3)(s_1+s_3)+2vt(1-s_1^2)(1-s_3^2)]}{(s_1+s_3 +2vt(1+s_1 s_3) )^3} 
\nonumber \\
 \times \exp \left\{\frac{M^2}{4v\Lambda^2}\frac{2s_1s_3+vt(s_1+ s_3 )}{(s_1+s_3 +2vt(1+s_1 s_3))}\right\},
 \label{asymp-pi}
\end{eqnarray}
where we have restored dimension of the light quark  and $\pi$-meson  masses, $m$ and $M_\pi$ respectively. Constant $f_\pi$  defined by Eq.~\eqref{asymp-pi}  exactly coincides  with the contribution of the diagram (a) in Fig.~\ref{fP_weak_decay_diagram} to the charged  pion weak decay constant, as it can be  seen from  Eq.~(27) in paper~\cite{Nedelko:2016gdk}. The contribution of the second diagram $(b)$ typical for all nonlocal approaches to the quark-meson vertices does not appear in the large $Q^2$ limit of the pion transition form factor.


\begin{thebibliography}{99}
\bibitem{Aubert:2009mc} 
  B.~Aubert {\it et al.} [BaBar Collaboration],
  Phys.\ Rev.\ D {\bf 80}, 052002 (2009)
  [arXiv:0905.4778 [hep-ex]].


\bibitem{Lepage:1980fj} 
  G.~P.~Lepage and S.~J.~Brodsky,
  Phys.\ Rev.\ D {\bf 22}, 2157 (1980).

\bibitem{Uehara:2012ag} 
  S.~Uehara {\it et al.} [Belle Collaboration],
  Phys.\ Rev.\ D {\bf 86}, 092007 (2012)
  [arXiv:1205.3249 [hep-ex]].

\bibitem{Mikhailov:2009kf} 
S.~V.~Mikhailov and N.~G.~Stefanis,
  Nucl.\ Phys.\ B {\bf 821}, 291 (2009)
  [arXiv:0905.4004 [hep-ph]].

\bibitem{Agaev:2010aq} 
  S.~S.~Agaev, V.~M.~Braun, N.~Offen and F.~A.~Porkert,
  Phys.\ Rev.\ D {\bf 83}, 054020 (2011)
  [arXiv:1012.4671 [hep-ph]].

\bibitem{Bakulev:2012nh} 
  A.~P.~Bakulev, S.~V.~Mikhailov, A.~V.~Pimikov and N.~G.~Stefanis,
  Phys.\ Rev.\ D {\bf 86}, 031501 (2012)
  [arXiv:1205.3770 [hep-ph]].

\bibitem{Klopot:2012hd} 
  Y.~Klopot, A.~Oganesian and O.~Teryaev,
  Phys.\ Rev.\ D {\bf 87} (2013) 036013 
  Erratum: [Phys.\ Rev.\ D {\bf 88}, no. 5, 059902 (2013)]
  [arXiv:1211.0874 [hep-ph]].

\bibitem{Oganesian:2015ucv} 
  A.~G.~Oganesian, A.~V.~Pimikov, N.~G.~Stefanis and O.~V.~Teryaev,
  Phys.\ Rev.\ D {\bf 93} (2016) 054040
  [arXiv:1512.02556 [hep-ph]].

\bibitem{Lucha:2011if} 
  W.~Lucha and D.~Melikhov,
  J.\ Phys.\ G {\bf 39}, 045003 (2012)
  [arXiv:1110.2080 [hep-ph]].

\bibitem{Li:2009pr} 
  H.~n.~Li and S.~Mishima,
  Phys.\ Rev.\ D {\bf 80}, 074024 (2009)
  [arXiv:0907.0166 [hep-ph]].

\bibitem{Kroll:2010bf}
  P.~Kroll,
  Eur.\ Phys.\ J.\ C {\bf 71} (2011) 1623
  [arXiv:1012.3542 [hep-ph]].

\bibitem{Gorchtein:2011vf} 
  M.~Gorchtein, P.~Guo and A.~P.~Szczepaniak,
  Phys.\ Rev.\ C {\bf 86}, 015205 (2012)
  [arXiv:1102.5558 [nucl-th]].

\bibitem{Zuo:2009hz} 
  F.~Zuo, Y.~Jia and T.~Huang,
  Eur.\ Phys.\ J.\ C {\bf 67}, 253 (2010)
  [arXiv:0910.3990 [hep-ph]].

\bibitem{Stoffers:2011xe} 
  A.~Stoffers and I.~Zahed,
  Phys.\ Rev.\ C {\bf 84}, 025202 (2011)
  [arXiv:1104.2081 [hep-ph]].

\bibitem{Swarnkar:2015osa} 
  R.~Swarnkar and D.~Chakrabarti,
  Phys.\ Rev.\ D {\bf 92}, no. 7, 074023 (2015) [arXiv:1507.01568 [hep-ph]].

\bibitem{Brodsky:2011yv} 
  S.~J.~Brodsky, F.~G.~Cao and G.~F.~de Teramond,
  Phys.\ Rev.\ D {\bf 84}, 033001 (2011)
  [arXiv:1104.3364 [hep-ph]].

\bibitem{Roberts:2010rn} 
  H.~L.~L.~Roberts, C.~D.~Roberts, A.~Bashir, L.~X.~Gutierrez-Guerrero and P.~C.~Tandy,
  Phys.\ Rev.\ C {\bf 82}, 065202 (2010)
  [arXiv:1009.0067 [nucl-th]].

\bibitem{Dorokhov:2010bz} 
  A.~E.~Dorokhov,
  arXiv:1003.4693 [hep-ph].

\bibitem{Dorokhov:2010zzb} 
  A.~E.~Dorokhov,
  JETP Lett.\  {\bf 92}, 707 (2010).
  
\bibitem{Dorokhov:2013xpa} 
  A.~E.~Dorokhov and E.~A.~Kuraev,
  Phys.\ Rev.\ D {\bf 88}, no. 1, 014038 (2013)
  [arXiv:1305.0888 [hep-ph]].

\bibitem{Kotko:2009mb} 
  P.~Kotko and M.~Praszalowicz,
  Phys.\ Rev.\ D {\bf 81}, 034019 (2010)
  [arXiv:0912.0029 [hep-ph]].
\bibitem{Lih:2012yu} 
  C.~C.~Lih and C.~Q.~Geng,
  Phys.\ Rev.\ C {\bf 85}, 018201 (2012)
  [arXiv:1201.2220 [hep-ph]].

\bibitem{Lichard:2010ap} 
  P.~Lichard,
  Phys.\ Rev.\ D {\bf 83}, 037503 (2011)
  [arXiv:1012.5634 [hep-ph]].

\bibitem{Arriola:2010aq} 
  E.~Ruiz Arriola and W.~Broniowski,
  Phys.\ Rev.\ D {\bf 81}, 094021 (2010)
  [arXiv:1004.0837 [hep-ph]].


\bibitem{Czyz:2012nq} 
  H.~Czyz, S.~Ivashyn, A.~Korchin and O.~Shekhovtsova,
  Phys.\ Rev.\ D {\bf 85}, 094010 (2012)
  [arXiv:1202.1171 [hep-ph]].


\bibitem{Kochelev:2009nz} 
  N.~I.~Kochelev and V.~Vento,
  Phys.\ Rev.\ D {\bf 81}, 034009 (2010)
  [arXiv:0912.2172 [hep-ph]].


\bibitem{McKeen:2011aa} 
  D.~McKeen, M.~Pospelov and J.~M.~Roney,
  Phys.\ Rev.\ D {\bf 85}, 053002 (2012)
  [arXiv:1112.2207 [hep-ph]].


\bibitem{Pham:2011zi} 
  T.~N.~Pham and X.~Y.~Pham,
  Int.\ J.\ Mod.\ Phys.\ A {\bf 26}, 4125 (2011)
  [arXiv:1101.3177 [hep-ph]].

\bibitem{Dorokhov:2009dg} 
  A.~E.~Dorokhov,
  Phys.\ Part.\ Nucl.\ Lett.\  {\bf 7}, 229 (2010)
  [arXiv:0905.4577 [hep-ph]].
    
\bibitem{Radyushkin:2009zg} 
  A.~V.~Radyushkin,
  Phys.\ Rev.\ D {\bf 80}, 094009 (2009)
  [arXiv:0906.0323 [hep-ph]].

\bibitem{Polyakov:2009je} 
  M.~V.~Polyakov,
  JETP Lett.\  {\bf 90}, 228 (2009)
  [arXiv:0906.0538 [hep-ph]].

\bibitem{EN1}   G.V. Efimov, and S.N. Nedelko,  Phys. Rev. D \textbf{51}, 176 (1995).


\bibitem{EN}  J.~V.~Burdanov, G.~V.~Efimov, S.~N.~Nedelko, S.~A.~Solunin,
 Phys. Rev. D\textbf{ 54}, 4483 (1996).

\bibitem{NK1}   A.C. Kalloniatis and S.N. Nedelko,  Phys. Rev. D \textbf{64}, 114025  (2001);

\bibitem{NK4}   
 A.C. Kalloniatis and S.N. Nedelko,  Phys. Rev. D \textbf{69}, 074029 (2004).

\bibitem{Nedelko:2014sla}
  S.~N.~Nedelko and V.~E.~Voronin,
  Eur.\ Phys.\ J.\ A {\bf 51} (2015) no.4,  45
  [arXiv:1403.0415 [hep-ph]].
  
\bibitem{Nedelko:2016gdk} 
  S.~N.~Nedelko and V.~E.~Voronin,
  Phys.\ Rev.\ D {\bf 93}, 094010 (2016)
  [arXiv:1603.01447 [hep-ph]].

\bibitem{Radyushkin:1996tb} 
  A.~V.~Radyushkin and R.~T.~Ruskov,
  Nucl.\ Phys.\ B {\bf 481}, 625 (1996)
  [hep-ph/9603408].

\bibitem{Bernard:1993wf}
  V.~Bernard, A.~H.~Blin, B.~Hiller, U.~G.~Meissner and M.~C.~Ruivo,
  Phys.\ Lett.\ B {\bf 305} (1993) 163
  [hep-ph/9302245].
 
\bibitem{Deng:2013uca}
  H.~B.~Deng, X.~L.~Chen and W.~Z.~Deng,
  Chin.\ Phys.\ C {\bf 38} (2014) no.1,  013103
  [arXiv:1304.5279 [hep-ph]].

\bibitem{PDG} K.~A.~Olive {\it et al.} [Particle Data Group Collaboration],
  Chin.\ Phys.\ C {\bf 38}, 090001 (2014).

\bibitem{Chiu:2007bc} 
  T.~W.~Chiu {\it et al.}  [TWQCD Collaboration],
  PoS LAT {\bf 2006}, 180 (2007).

\bibitem{Behrend:1990sr} 
  H.~J.~Behrend {\it et al.} [CELLO Collaboration],
  Z.\ Phys.\ C {\bf 49}, 401 (1991).

\bibitem{Gronberg:1997fj} 
  J.~Gronberg {\it et al.} [CLEO Collaboration],
  Phys.\ Rev.\ D {\bf 57}, 33 (1998)
  [hep-ex/9707031].



\bibitem{BABAR:2011ad} 
  P.~del Amo Sanchez {\it et al.} [BaBar Collaboration],
  Phys.\ Rev.\ D {\bf 84}, 052001 (2011)
  [arXiv:1101.1142 [hep-ex]].

\bibitem{Lees:2010de} 
  J.~P.~Lees {\it et al.} [BaBar Collaboration],
  Phys.\ Rev.\ D {\bf 81}, 052010 (2010)
  [arXiv:1002.3000 [hep-ex]].

\bibitem{Kalloniatis:2002ct} 
  A.~C.~Kalloniatis and S.~N.~Nedelko,
  Phys.\ Rev.\ D {\bf 66}, 074020 (2002)
  [hep-ph/0208064].

\end{thebibliography}
\end{document}